\documentclass[12pt]{iopart}

\usepackage{iopams}
\usepackage{graphicx}
\expandafter\let\csname equation*\endcsname\relax
\expandafter\let\csname endequation*\endcsname\relax
\usepackage{amsmath}
\usepackage{color}

\newcommand{\blue}[1]{\textcolor{black}{#1}}

\begin{document}

\title[$\Phi_1$ in electron root stellarator plasmas]{On-surface potential and radial electric field variations
in electron root stellarator plasmas}


\author{J. M. Garc\'ia-Rega\~na$^1$, T. Estrada$^1$, I. Calvo$^1$, J. L. Velasco$^1$,
J. A. Alonso$^1$, D. Carralero$^1$, R. Kleiber$^2$, M. Landreman$^3$, A. Moll\'en$^2$, 
E. S\'anchez$^1$, C. Slaby$^2$, TJ-II Team and W7-X Team}
\address{$^1$ Laboratorio Nacional de Fusi\'on CIEMAT, Av. Complutense 40, 28040 Madrid, Spain\\
$^2$ Max-Planck-Institut f\"ur Plasmaphysik, Wendelsteinstr. 1, 17491 Greifswald, Germany\\
$^3$ Institute for Research in Electronics and Applied Physics, University of Maryland, College Park, Maryland 20742, USA}
\ead{jose.regana@ciemat.es}
\vspace{10pt}

\begin{abstract}
In the present work we report recent radial electric field
measurements carried out with the Doppler reflectometry system
in the TJ-II stellarator.
The study focuses on the fact that, under some conditions, 
the radial electric field measured at different points over the same flux surface 
shows significantly different values. A numerical analysis
is carried out considering the contribution 
arising from the radial dependence of $\Phi_1$ as a possible correction term to 
the total radial electric field. Here $\Phi_1$
is the neoclassical electrostatic potential variation over the surface.
The comparison shows good agreement in some aspects, like the conditions under which 
this correction is large (electron-root conditions) or negligible (ion-root conditions). But it disagrees
in others like the sign of the correction. The results are discussed together with 
the underlying reasons of this partial disagreement.\\
In addition, motivated by the recent installation of the dual Doppler reflectometry 
system in Wendelstein 7-X (W7-X), $\Phi_1$ estimations for W7-X are revisited 
considering Core-Electron-Root-Plasma (CERC) plasmas from its first experimental campaign. 
The simulations show larger values of $\Phi_1$
under electron-root conditions than under ion root ones. 
The contribution from the kinetic electron response is shown to become
important at some radii. All this results in a potential variation size 
noticeably larger than estimated in our previous work in W7-X \cite{Regana_nf_57_056004_2017}
for other plasma parameters and another configuration.
\end{abstract}

%
%
%
%
%

\section{Introduction}

The radial electric field is one of the physical
quantities with significant prominence in stellarator transport physics problems.
In particular, for the radial transport of impurities and 
their accumulation, its role becomes more important as the
charge state of the impurity increases. In stellarators, the explanation for this is framed 
by the standard neoclassical formalism. 
There, one can express the flux-surfaced-averaged fluxes as a linear 
combination of thermodynamic forces 
and the so-called thermal transport matrix coefficients $L_{ij}^{a}$:

\begin{equation}
\frac{\left<\boldsymbol{\Gamma}_{a}\cdot\nabla r\right>}{n_{a}}=
-L_{11}^{a}\left(\frac{n_a'}{n_a}+\frac{Z_a e E_{r}}{T_{a}}+ \frac{L_{12}^{a}}{L_{11}^{a}}\frac{T_a'}{T_a}\right),
\label{eqn:flows}
\end{equation}
with $a$ the species index, $n_{a}$ the density, $T_{a}$ the temperature, 
$Z_{a}$ the charge state, $e$ is the proton charge, $\boldsymbol{\Gamma}_a$ the 
particle flux density and $\left< ... \right>$ the flux surface average operator.
In the present work $r$ is a flux surface label with the character of an
effective radial coordinate such that the volume enclosed by a flux surface is 
$V=2\pi^2 R_0 r^2$, and $R_{0}$ is the major radius of the stellarator. The prime $'$ denotes differentiation with respect to $r$.
The radial electric field vector is $\mathbf{E}_{r}=E_{r}\nabla r$ with 
$E_{r}=-\Phi_0'$ and $\Phi_0=\Phi_0(r)$ the part of the electrostatic potential constant on the flux surface.
An important well-known difference between particle transport in stellarators respect to that in 
(axi-symmetric) tokamaks is that the particle transport of the different species does not obey
ambipolarity at any $E_{r}$. In other words, the total radial flux-surface-averaged current does not vanish
and quasi-neutrality is not preserved along the radial direction. 
Then, the radial electric field in stellarators is determined by imposing this ambipolarity condition, that reads, 
see e.g. \cite{calvo_ppcf_55_125014_2013},

\begin{equation}
\sum_{a}Z_{a}e\left<\boldsymbol{\Gamma}_{a}\cdot\nabla r\right>=0.
\label{eqn:ambipolarity}
\end{equation}
Despite the apparently explicit linear dependence of the fluxes on $E_{r}$, see eq.~(\ref{eqn:flows}), 
the role that the radial electric field plays on the confinement of the trapped particle orbits
in the long-mean-free-path regimes, makes the matrix transport coefficients depend 
also on $E_{r}$. This leads the ambipolarity condition to become a non-linear equation with multiple 
roots \cite{Mynick_nf_23__1983, Hastings_nf_25_4_1985}. However, only two of them are usually identified
in laboratory plasmas.
For simplicity, assuming the presence of only bulk ions and electrons, 
when the collisionality of both species is such that
the radial particle flux of the ions needs to be reduced in order to 
satisfy ambipolarity, the ambipolar electric field typically points 
radially inward and
$E_{r}<0$. If, on the contrary, the electron radial particle flux needs 
to be retarded to fulfill ambipolarity the radial electric field points radially outwards, $E_{r}>0$.
These two situations are referred to as ion and electron root regimes respectively.
In general, standard neoclassical theory predicts ion root 
conditions for all collisionalities when the
ion and electron temperatures are comparable, $T_{i}\sim T_{e}$, and fairly large and positive (electron root) $E_{r}$ values
at low collisionality with strongly localized electron heating that leads to $T_{e}\gg T_i$,
see e.g. \cite{Maassberg_pop_5_1993, Maassberg_ppcf_41_1999}.
The concern for the intrinsic character of the impurity 
accumulation in stellarators and ion root conditions has been traditionally tight together, since
the inward pinch related to $E_{r}$ can, for sufficiently 
high $Z_{a}$, exceed in most situations the outward counterparts driven by 
the temperature and density gradients. This has also been observed in numerous stellarator experiments, 
see e.g. \cite{Burhenn_nf_49_2009} and references therein.\\

However, this simple explanation concerning impurity accumulation has been broadened in recent 
years motivated by a few experiments that question it, like the hollow impurity 
density profiles observed in LHD \cite{Ida_pop_16_056111_2009} or the exceptionally low
impurity confinement time of the HDH mode in
W7-AS plasmas \cite{McCormick_prl_89_015001_2002}. 
For decades it has been known that a variation of the electrostatic potential over the flux surface 
$\Phi_1=\Phi_1(r,\theta,\phi)$ can be relatively large for low collisionality plasmas in non-omnigeneous stellarators
\cite{Mynick_nf_23__1983, Ho_pf_30.2_1987, Beidler_isw_2005}. Here $\theta$ and $\phi$
are some angular poloidal and toroidal coordinates, respectively.
This piece of the 
electrostatic potential is necessary 
in order to restore quasi-neutrality over the flux surfaces, which the cumulative effect of the non-vanishing bounce-averaged radial 
displacement of the particle drift orbits violates. 
Although it can be in most situations negligible for main ion and electron transport, the importance of $\Phi_1$ for impurities resides in the fact 
that the radial component of the $E\times B$ drift, $\mathbf{v}_{E1}=-\nabla\Phi_1\times \mathbf{B}/B^2$, can become of the order of the
radial component of the magnetic drift $\mathbf{v}_{m}$, basically because the latter scales as $Z_{a}^{-1}$ while 
the former does not. 
Consequently, its role as source of radial transport can become as important as the
inhomogeneous confining magnetic field for sufficiently high charge state.
Since the first numerical calculations of $\Phi_1$ \cite{Garcia-Regana2013} 
performed with the code EUTERPE \cite{Kornilov_pop_11_2004, Kornilov_nf_45.4_2005} and the experimental measurement in 
a stellarator \cite{Pedrosa_nf_55_2015}, other works have followed this line:
the estimation of its effect on the radial flux of impurities for some selected ion-root 
plasmas for different stellarator configurations in ref.~\cite{Regana_nf_57_056004_2017}; 
the analytical development of the formalism \cite{Calvo_ppcf_59_055014_2017} and 
the code (KNOSOS) \cite{Velasco_ppcf_accepted_2018}
that integrates the drift kinetic equation and transport quantities of interest, 
including $\Phi_1$, for optimized stellarators; new LHD impurity plasmas analyzed under the effect of $\Phi_1$
with the SFINCS code \cite{Landreman_pop_21_042503_2014, sfincs_github}, 
including the self-consistent modification of $E_{r}$ by $\Phi_1$ 
and including non-tracer impurities \cite{Mollen_submitted_2018}. Apart from these works,
others have looked into the screening of impurities
in stellarators, like ref.~\cite{Velasco_nf_57_1_2017} where high $T_{i}$ plasmas with negative 
but small $|E_{r}|$ are shown to coexist with outward impurity flow. 
Finally, 
ref.~\cite{Helander_prl_118_155002_2017} studies analytically the radial particle flux of 
highly collisional impurities in low collisional bulk plasmas, concluding that in the 
case without $\Phi_1$, the
radial transport of impurities may only weakly depend on $E_{r}$ and temperature screening can arise;
and ref.~\cite{Calvo_prl_submitted_2018} where the previous derivation is generalized including $\Phi_1$, 
which makes the impurity radial particle flux to depend strongly on $E_r$.\\

The conclusions from the works dealing with $\Phi_1$ \cite{Regana_nf_57_056004_2017, 
Velasco_ppcf_accepted_2018, Mollen_submitted_2018} coincide on their prediction about its size, 
that reaches for LHD values of up to $e\Delta\Phi_1/T_{i}\sim 0.1$, with 
$\Delta \Phi_1=(\Phi_1^{\text{max}}-\Phi_1^{\text{min}})$ and $\Phi_1^{\text{max}}$ and $\Phi_1^{\text{min}}$ its maximum 
and minimum value respectively over a given flux surface. 
The direction and magnitude of the impact of $\Phi_1$ on the impurity radial transport is not trivial.
It depends on the charge state of the impurities, the collisional regime where the impurity is, how its distribution function 
couples to $\Phi_1$, etc. However, based on 
the available numerical simulations it can be stated without too much lack of generality that variations of that size 
undeniably introduces a strong correction to the standard neoclassical prediction in LHD, even
considering low-$Z$ impurities like carbon. For TJ-II similar values of the normalized 
potential variation are also predicted \cite{Regana_nf_57_056004_2017}, even at the higher collisionality of 
its plasmas. Moreover, the estimations in TJ-II qualitatively agree with the experimental measurements of the plasma floating 
potential difference at the edge flux surfaces \cite{Pedrosa_nf_55_2015}. 
Regarding W7-X, the variations are typically shown one order of magnitude lower than those for LHD plasmas at comparable
collisionality. However, as noted below only few simulations are available, in particular,
for the magnetic configurations and plasma parameters from the experimental campaigns.\\

The present work aims at broadening the scanned parameter space with a comparative 
view between ion root and electron root conditions in TJ-II and W7-X, with the focus
mainly on the second of these regimes. There are several reasons for this: 
first, all the numerical effort has looked so far into ion root plasmas, with the 
underlying hope that $\Phi_1$ could, at least, cancel the predicted $E_r$-driven inward pinch.
A similar analysis for electron-root plasmas is missing despite the fact that 
$\Phi_1$ can indeed be larger than in ion-root for the same absolute value
of $E_{r}$, as pointed out in \cite{Pedrosa_nf_55_2015}; 
second, although its impact is predicted to be large for impurities, the size of $\Phi_1$ 
is still small compared to the lower order part of the potential $\Phi_0$, 
and its direct detection is instrumentally difficult. In the present work, 
under the light of recent Doppler reflectometry (DR) measurements of the radial
electric field in TJ-II, where strong differences over the same flux surfaces have
been found under electron
root conditions, we investigate whether the radial dependence of the calculated 
$\Phi_1$ can explain those differences; and finally,
since the configuration and parameter space of W7-X is rather large \cite{Geiger_ppcf_57_1_2015}, 
the results obtained for the few configurations and parameters considered in 
\cite{Regana_nf_57_056004_2017, Mollen_submitted_2018} should not be generalized. 
In the present work,
we have based our calculations in typical parameters of
OP1.1 \cite{Klinger_ppcf_59_2017} Core-Electron-Root-Confinement (CERC) plasmas considering a configuration 
with large effective ripple. We show numerically that $\Phi_1$ can then be as large as in the reported LHD cases. 
This exercise has also been performed considering adiabatic and kinetic electrons, in 
order to provide explicitly a validity check for the adiabatic electron approximation, 
that for codes like EUTERPE can result in considerably less computational time.\\

After this section, a brief overview of the equations and tools employed 
are described in section \ref{sec:eqs}. The TJ-II results, both numerical and experimental, 
are presented and discussed in section \ref{sec:tj2}. The numerical analysis for W7-X
CERC conditions is shown in section \ref{sec:w7xop11}. Finally the conclusions are
summarized in section \ref{sec:conclusions}.

 

\section{Equations and numerical methods}
\label{sec:eqs}

In this section we give an overview of the numerical method, 
the relevant equations of the problem and the numerical code used, EUTERPE \cite{Kornilov_pop_11_2004, Kornilov_nf_45.4_2005}. 
The content of this section concerns the neoclassical version of the code. 
For a more complete description of how the present problem is approached we refer the reader to section 2 of ref.~\cite{Regana_nf_57_056004_2017}. 
Other aspects dealing with the neoclassical version can be found in 
refs.~\cite{Kauffmann_jpcs_260.1_2010, Garcia-Regana2013, Kauffmann_thesis}, and those closer to
the numerical implementation in refs.~\cite{Borchardt_jcp_231_2012, Kleiber_cpc_183_2012, Kleiber_cpc_182_1005_2011}.\\

EUTERPE is a $\delta f$ particle-in-cell (PIC) Monte Carlo
code. For a given kinetic species, it considers a splitting of the
distribution $f=f_{0}+ f_{1}$, with $f_{0}$ an analytically 
known expression with the role of a control variate, which does not have to be necessarily 
linked to any approximation. The code solves the kinetic equation 
for the $f_{1}$, $\mathrm{d}f_{1}/\mathrm{d}t=-\mathrm{d} f_0/\mathrm{d}t + C(f)$, with 
$C(f)$ a collision operator. 
The choice of phase space coordinates is the following: 
in real space, in order to characterize the guiding center position 
$\mathbf{R}$ of the Monte Carlo markers,  
the magnetic PEST \cite{Grimm_jcp_49_1983} poloidal and toroidal angles $\theta$
and $\phi$, and a flux surface label $r$ are employed. In velocity space
the parallel component of the velocity $v_{\|}$ and 
normalized magnetic moment $\mu=v_{\bot}^{2}/2B$ are considered. 
Here $f_0=f_{M}\exp(-Ze\Phi_1/T)$, with $f_{M}=\left[n_{0}/(2\pi)^{3/2} 
v_{th}^{3}\right]\exp\left[-\left(v_{\|}^{2}+v_{\bot}^2\right)/v_{th}^{2}\right]$ 
the Maxwellian distribution, $v_{\bot}$ the
perpendicular component of the velocity, $n_{0}=n_{0}(r)$ the density, $T$ the temperature, 
$v_{th}=\sqrt{2T/m}$ the thermal speed, $m$ the mass and $B$ the magnetic field strength.
With these definitions, the kinetic equation takes the form:

\begin{equation}
\label{eq:fokker-planck}
\frac{\partial f_1}{\partial t}+
\dot{\mathbf{R}}\cdot\nabla f_1+
\dot{v}_{\|}\frac{\partial f_1}{\partial v_{\|}}=
-f_{\text{M}}\left(\mathbf{v}_{m}+\mathbf{v}_{E1}\right)\cdot\nabla r
\left[\frac{n'}{n}+\frac{Ze}{T}\Phi_0'+
\left(\frac{mv^{2}}{2T}-\frac{3}{2}+\frac{Ze}{T}\Phi_1\right)\frac{T'}{T}\right]+C(f).
\end{equation}
The overdot $\dot{}$ denotes differentiation with respect to time $t$.
Finally, the following equations of motion
enter the left-hand-side of eq.~(\ref{eq:fokker-planck}):

\begin{eqnarray}
  &&\dot{\bf R} = v_\| {\bf b} + \frac{{\bf b} \times \nabla \Phi_0}{B},
  \label{eq:dotR_sim}\\
  &&\dot v_\| = - \frac{\mu}{m_a} {\bf b} \cdot \nabla B - \frac{v_\|}{B^2} ({\bf b} \times {\nabla B} ) \cdot \nabla \Phi_0 - \frac{Z_a e}{m_a} {\bf b} \cdot \nabla \Phi_1,
  \label{eq:dotvpar_sim},\\
  &&\dot \mu = 0,  \label{eq:dotmu_sim}
\end{eqnarray}
%
with $\mathbf{b}=\mathbf{B}/B$.
In order to obtain $\Phi_1$, quasi-neutrality among all the species is imposed up to 
first order: $\sum_{a}Z_{a}en_{a}=0$, with 
$n_{a}=n_{0a}(r)\exp(-Z_{a}e\Phi_{1}/T_{a})+n_{1a}$
the density of the different species. Considering singly charged bulk ions (i) and electrons (e) and 
assuming $e\Phi_1/T\ll 1$, quasi-neutrality yields:

\begin{equation}
\label{eq:qngeneral}
\Phi_{1}=\frac{T_{e}}{e}
\left(n_{0e}+n_{0i}\frac{T_{e}}{T_{i}}\right)^{-1}
\left(n_{1i}-n_{1e}\right).
\end{equation}
Note that in ref.~\cite{Regana_nf_57_056004_2017}
the assumption of adiabatic electrons, i.e. $n_{e}\approx n_{0e}(r)\exp(e\Phi_{1}/T_{e})$, implies that
on the right-hand-side of expression (\ref{eq:qngeneral}) only $n_{1i}$ appears. 
In the present work, in section \ref{sec:w7xop11}, 
this approximation is relaxed and the results with adiabatic and kinetic electrons are compared with each other.\\

Another difference between ref.~\cite{Regana_nf_57_056004_2017} and the present work
is the treatment of the collision operator $C(f)$. While in ref.~\cite{Regana_nf_57_056004_2017} pitch angle
scattering collisions without momentum conservation were applied, in the present work, 
a momentum-restoring 
field particle term similar to that implemented in other codes 
\cite{Satake_pfs_3_s1062_2008, Vernay_pop_17_122301_2010} is added. The detailed description for EUTERPE can be found 
in ref.~\cite{Slaby_NF_accepted_2018}.

\section{Potential variations in TJ-II: comparison between ion- and electron-root plasmas}
\label{sec:tj2}

\subsection{TJ-II Doppler reflectometry system}
\label{sec:instrum}

TJ-II is a heliac-type stellarator where, for the standard configuration 
considered for this work, the average magnetic field is $0.95$ T on axis, 
the rotational transform is $\iota\approx 1.5$ at the center of the plasma and $1.6$ approximately at the edge
and the effective minor radius and major radius are $a=0.2$ m and
$R_{0}=1.5$ m, respectively. The available heating power consists of two gyrotrons
delivering $300$ kW each (operated both in X-mode at the second harmonic of the electron cyclotron
frequency) and two NBI heating systems, one co- and another counter-injecting 
each a port-through power of up to 700 kW. For the results presented below, 
only ECH on-axis was used. With this heating scheme the central electron density 
typically reaches values of $n_{e}\approx 0.5 - 1\times 10^{19}$ m$^{-3}$, the electron temperature $T_{e}\approx 1-2$ keV and the ion temperature $T_{i}\approx 80-100$ eV.\\

\begin{figure}
  \centering
  \includegraphics[width=0.48\textwidth]{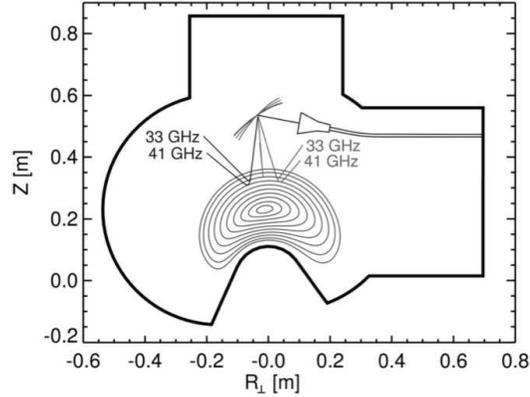}
  \caption{Schematic representation of the TJ-II vacuum vessel with DR antenna-mirror arrangement showing 
the two plasma regions that can be probed by the system. Here $R_{\bot}=R-R_0$, with $R$ the cylindrical radial 
coordinate used below and $R_{0}$ the major radius
of the device.}\label{fig:DRsystem}
\end{figure}

For the experimental results discussed in this section
the technique used has been Doppler reflectometry (DR). It allows the measurement of density fluctuations
and their perpendicular rotation velocity at different turbulence scales, with good spatial and
temporal resolution. From the perpendicular rotation velocity the radial electric field, the 
central quantity in this section, can be obtained. 
The DR in operation at TJ-II \cite{Happel_rsi_80_2009} works in a frequency
hopping mode in the Q-band: 33-50 GHz, covering typically the radial region from $r/a=0.6$ to
$r/a=0.9$. Its front-end consists of a compact corrugated antenna and an ellipsoidal mirror. The
mirror can be tilted to probe different perpendicular wave-numbers of the turbulence in the
range $k_{\bot}\approx 1-14$ cm$^{-1}$, at different plasma regions poloidally separated, as both positive and
negative probing beam angles with respect to normal incidence can be selected, see
fig. \ref{fig:DRsystem}. Assuming that the electron density 
is constant on each flux surface, this characteristics makes possible to access different 
points of measurement over the same flux surface. 
Apart from its interest for studying the spatial localization of instabilities 
predicted in stellarators by gyro-kinetic
simulations \cite{Kornilov_pop_11_2004, Xanthopoulos_prx_6_2016, Riemann_ppcf_58_2016}, 
for the results presented in this work this feature has been exploited to characterize the 
radial electric field measured on the left and right regions with respect to the incidence angle
where the launched beam is normal to the last closed flux surface. 
Throughout the present section these regions are referred to as ``left'' and ``right''
regions, see fig. \ref{fig:DRsystem}.

\subsection{Experimental and numerical results}
\label{sec:tjii_exp_results}

\begin{figure}
  \includegraphics[width=0.48\textwidth]{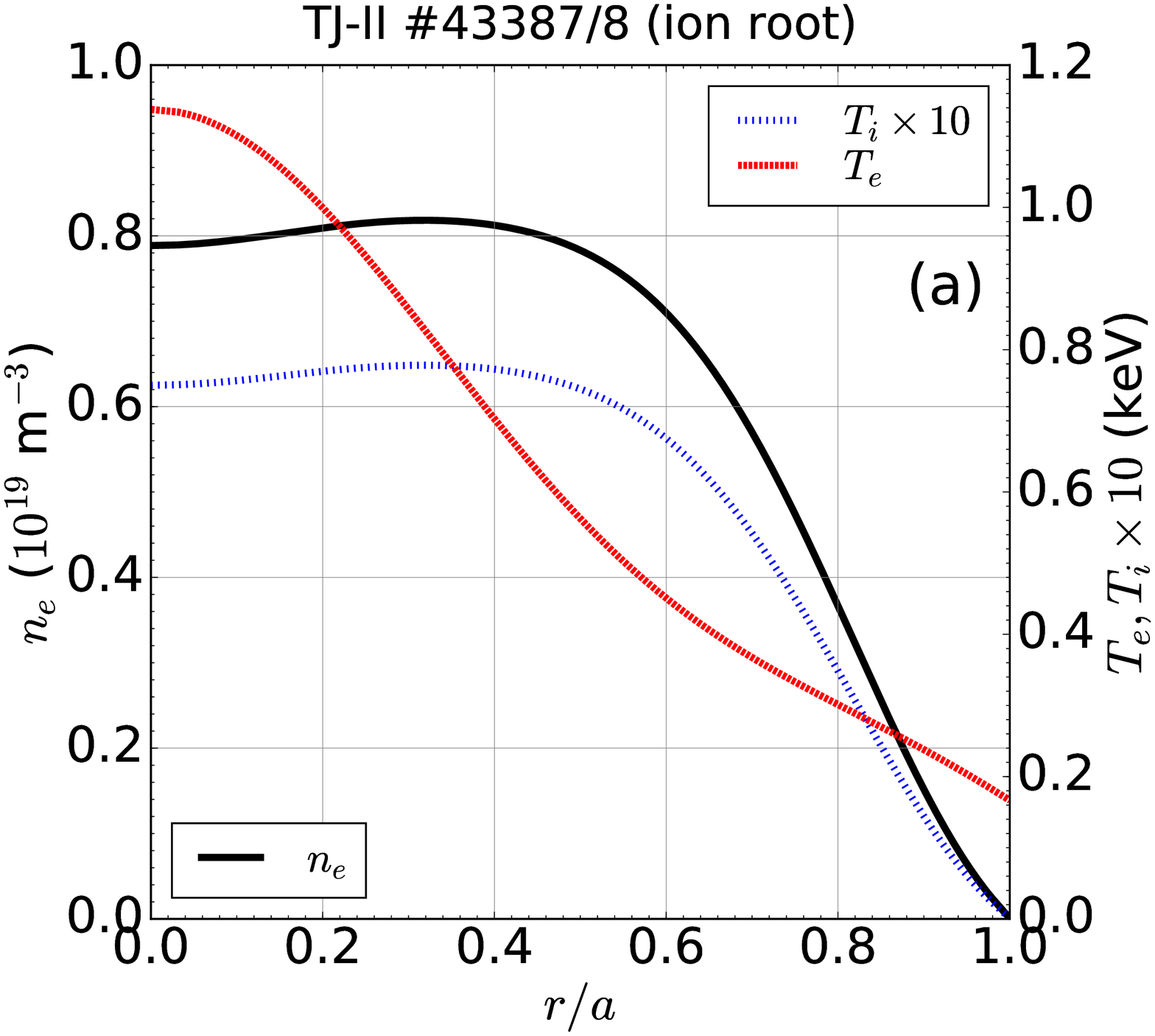}
  \includegraphics[width=0.48\textwidth]{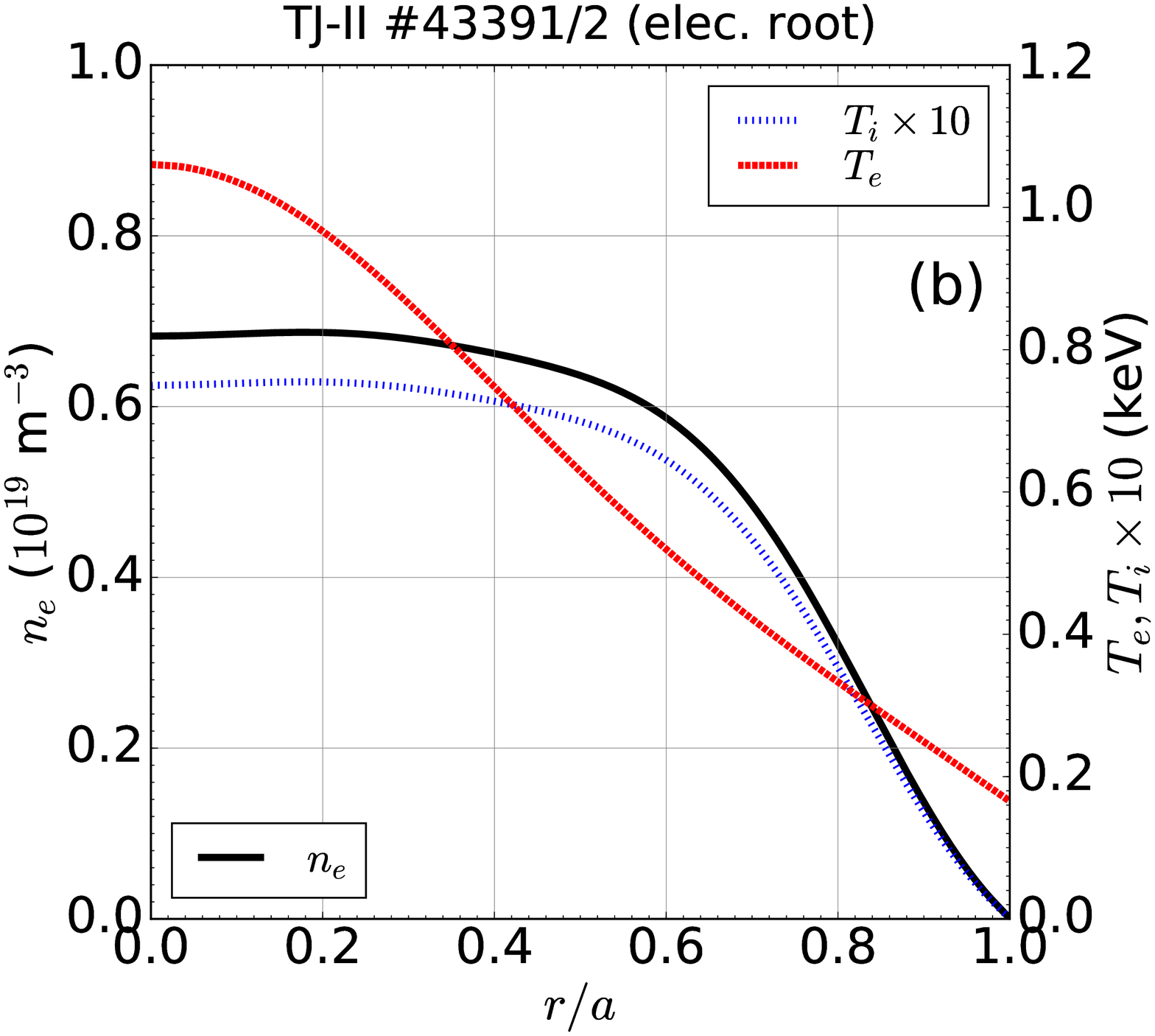}
  \caption{Left: radial profiles of electron density ($n_e$, solid black line), electron 
temperature ($T_{e}$, dashed red line) and ion temperature multiplied by $10$ ($T_{i}$, dotted 
blue line) considered for the EUTERPE simulations based on 
those of TJ-II discharges $\#43387$ and $\#43388$ measured with the
Thomson Scattering ($n_{e}$ and $T_{e}$) and the NPA ($T_{i}$) 
systems. Right: same quantities as on the left but considering the TJ-II discharges $\#43391$ and $\#43392$.}\label{fig:profs_pars}
\end{figure}

\begin{figure}
  \includegraphics[width=0.48\textwidth]{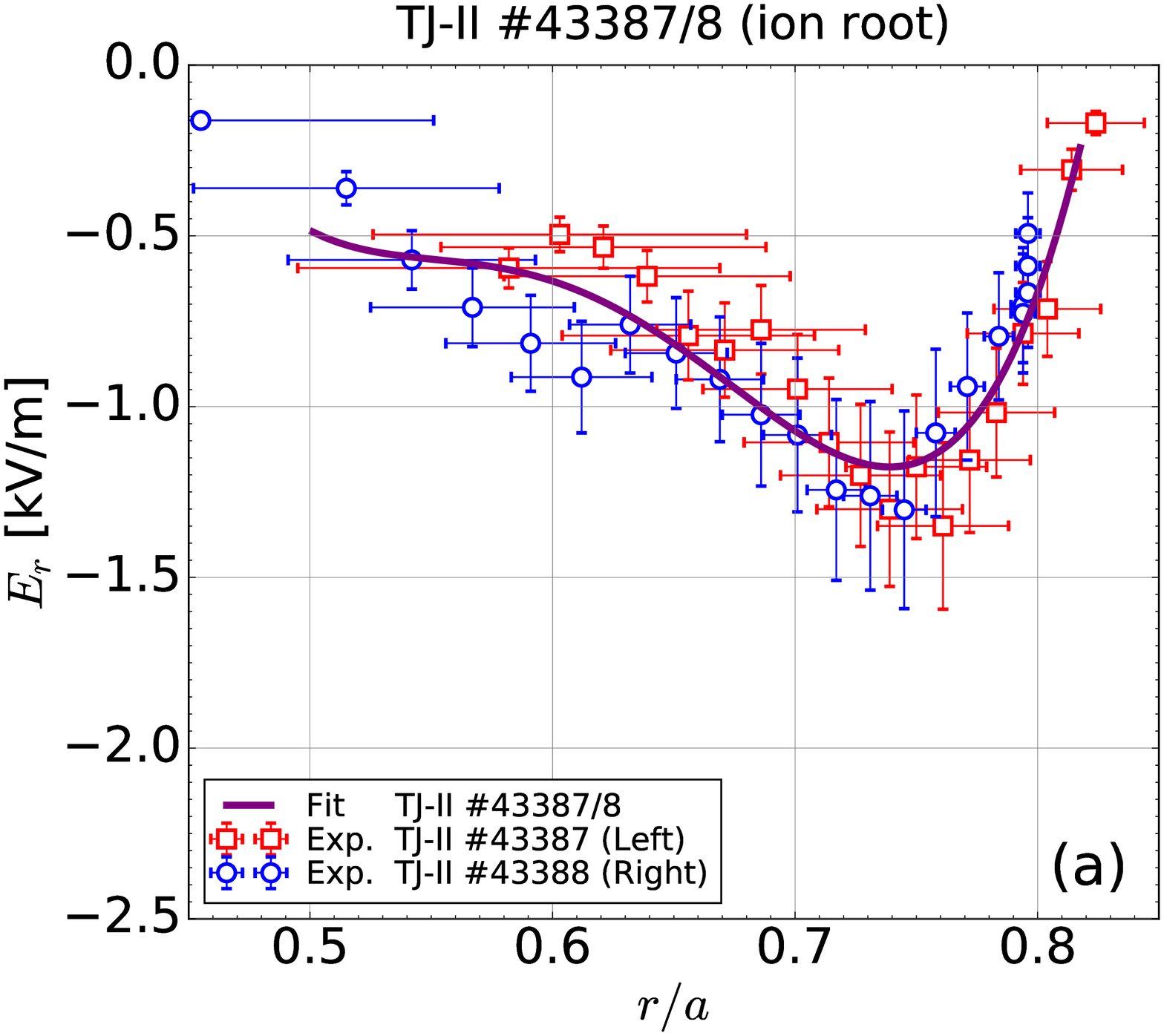}
  \includegraphics[width=0.48\textwidth]{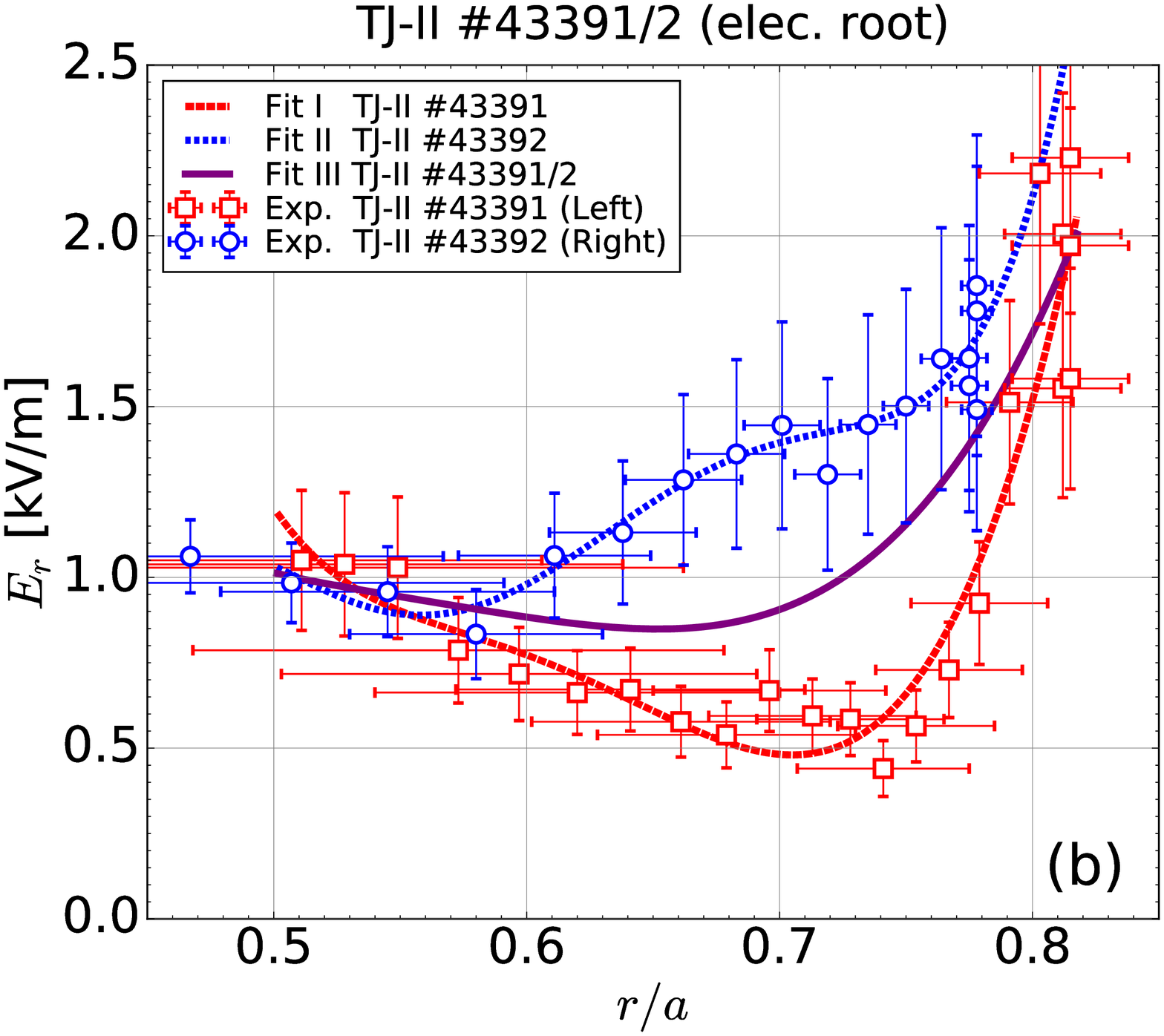}
  \caption{Left: radial electric field $E_{r}$ as a function of the normalized effective 
radius for the TJ-II ion root plasmas, discharges $\#43387$ and $\#43388$.
Right: the same but for the electron root plasmas from discharges $\#43391$ and $\#43392$. 
In both cases the measurements performed on the left and right sides of the 
DR measurement plane are represented with red squares and blue circles respectively. The solid lines
correspond to the input radial electric field profiles used for the $\Phi_1$ EUTERPE simulations.}
\label{fig:profs_er}
\end{figure}

Two pairs of TJ-II discharges are considered.
The main difference between them is the sign 
of the radial electric field. 
The first couple of discharges ($\#43387$ and $\#43388$)
are representative for ion root regime while the second couple 
of discharges ($\#43391$ and $\#43392$) are in electron root. 
The plasma parameters for each of these pairs are represented in 
figs. \ref{fig:profs_pars} (left) and \ref{fig:profs_pars} (right) respectively.
The difference in the density profiles determines what regime is accessed.
TJ-II plasmas exhibit this ion-to-electron root change when the 
line-averaged density, obtained with a microwave interferometer \cite{Sanchez_rsi_75_10_2004}, is close to the critical value of 
$\bar{n}_{e}^{\text{cr}}\sim 0.6\times 10^{19}$ m$^{-3}$ (for the standard magnetic configuration 
and the used heating power), which standard neoclassical
calculations
capture without difficulty, see e.g. \cite{Velasco_prl_106_135003_2012, Velasco_ppcf_55_12_2013}. 
The characteristic of the DR analysis that has motivated the numerical 
simulations is the difference that the radial electric field value, for each case, 
shows when the measurement is taken on the left probing region and on the right. 
This is equivalent to 
measuring different values of the radial electric field at different points 
over the same flux surface.\\

\begin{figure}
  \centering
  \includegraphics[height=4.1cm]{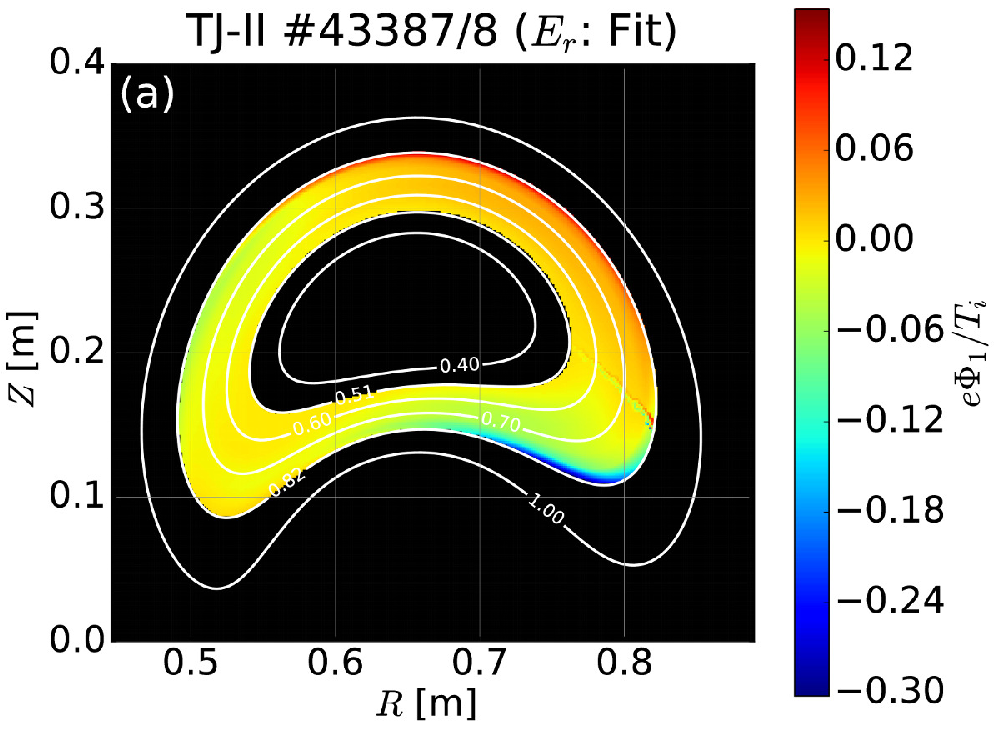}
  \includegraphics[height=4.1cm]{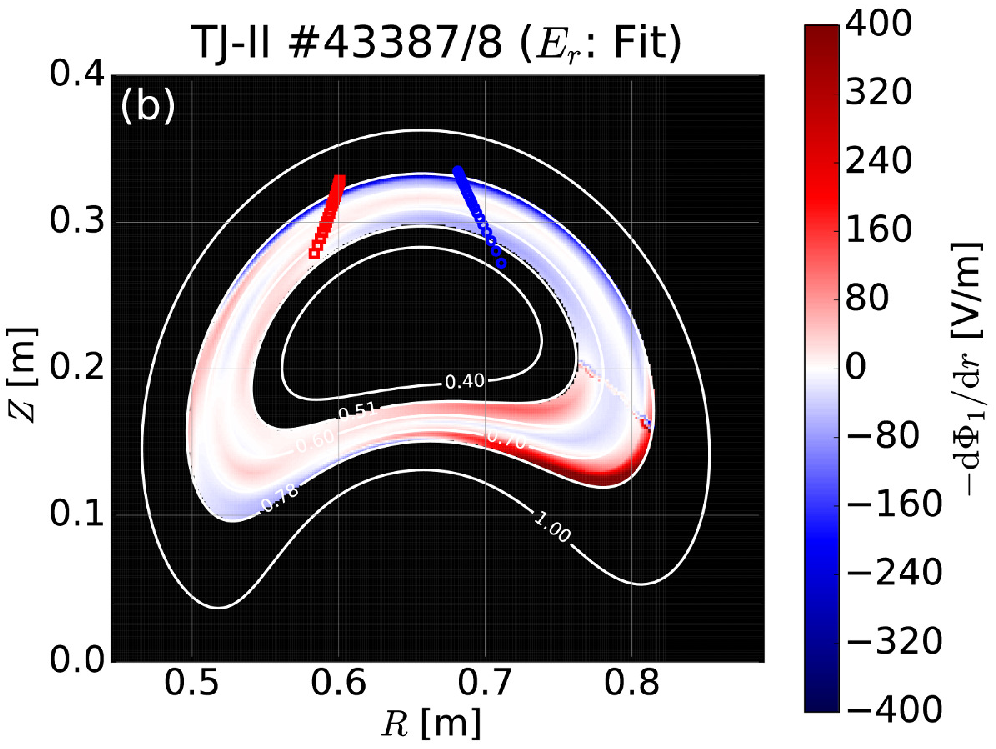}
  \includegraphics[height=4.1cm]{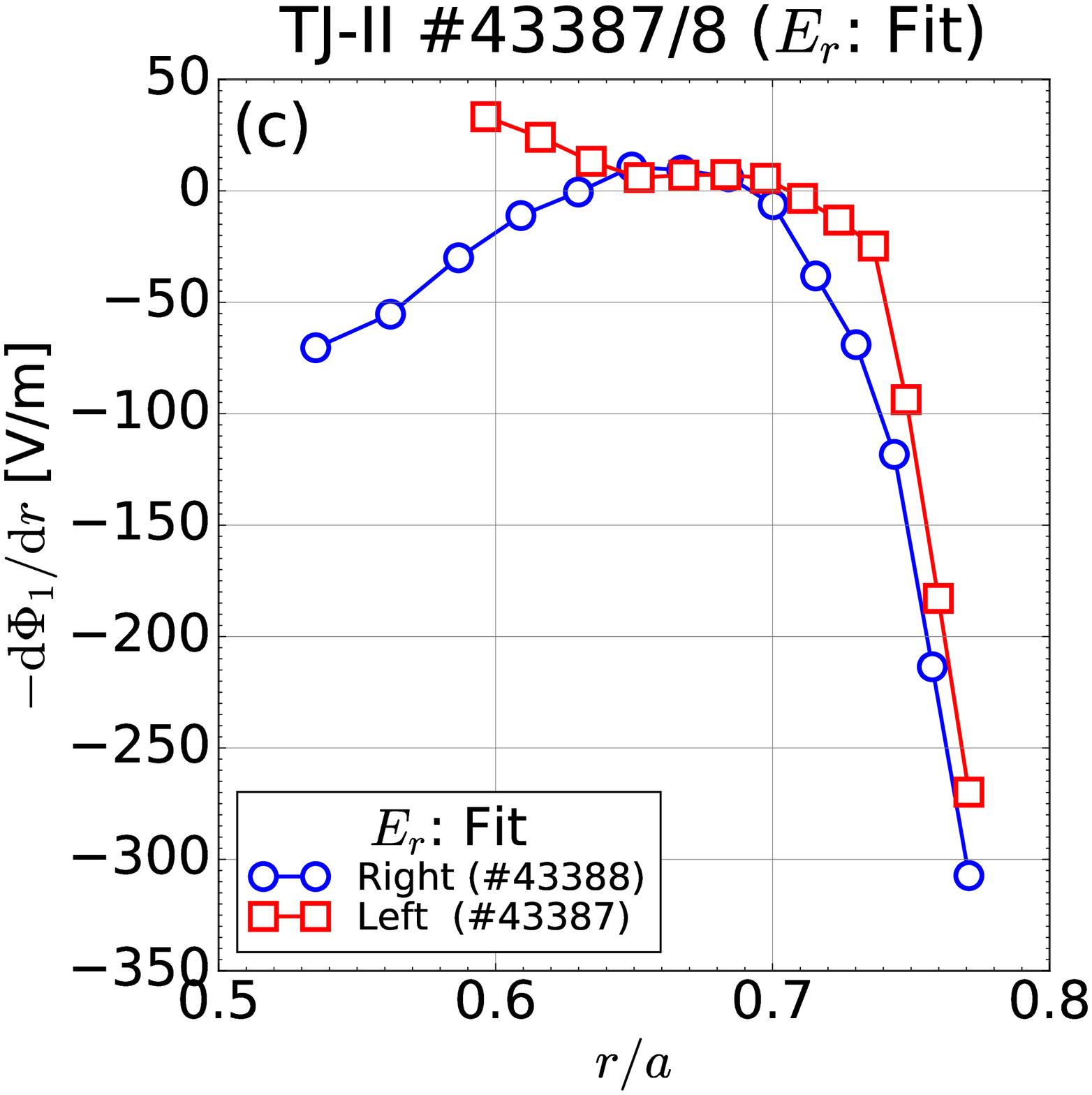}
  \caption{For the ion root conditions TJ-II plasmas: (a) potential variation normalized to the ion 
temperature $e\Phi_1/T_{i}$ at the Doppler reflectometry probing plane in the range of simulated 
radii; (b) Over the same plane, first order radial electric field $-\text{d}\Phi_1/\text{d} r$, together
with the specific positions of measurement on the left and right DR probing regions, 
estimated with ray tracing; (c) Value of 
$-\text{d}\Phi_1/\text{d} r$ at those positions where, as before, red squares and right blue circles correspond
to the estimations along the left and right measurement positions respectively.}\label{fig:tj2ion}
\end{figure}

\begin{figure}
  \centering
  \includegraphics[height=3.95cm]{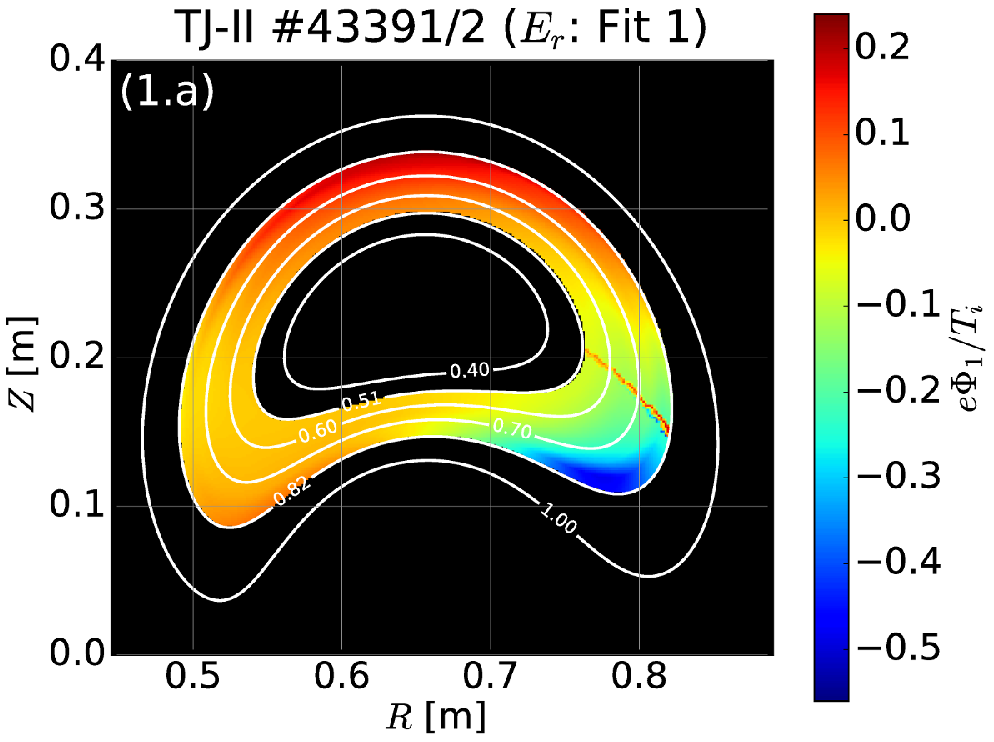}
  \includegraphics[height=3.95cm]{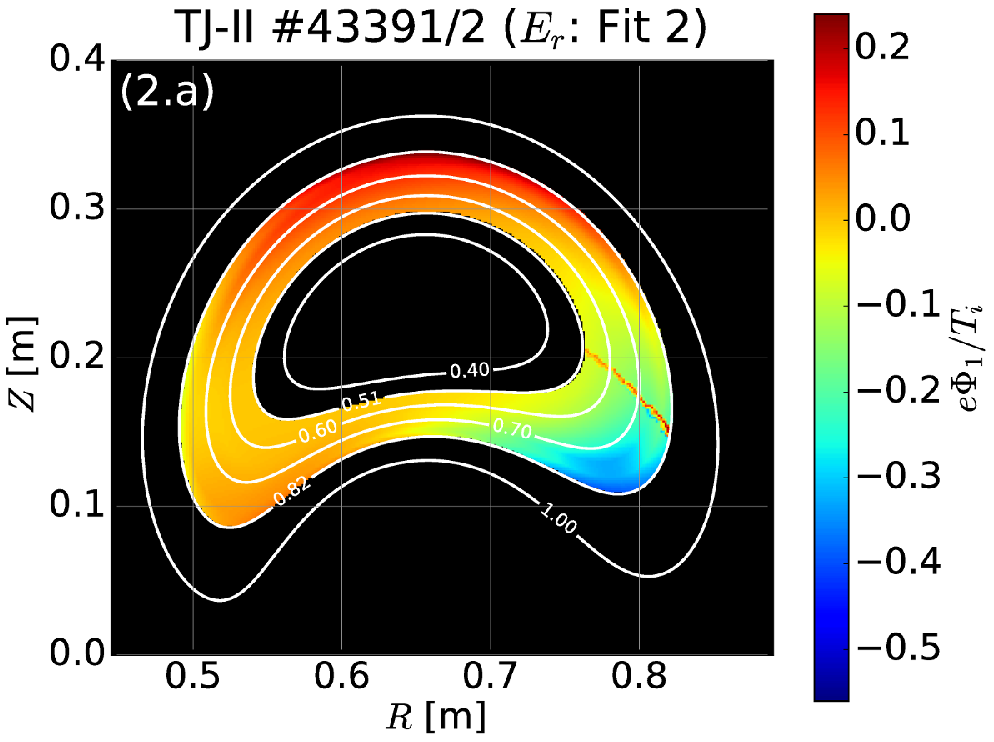}
  \includegraphics[height=3.95cm]{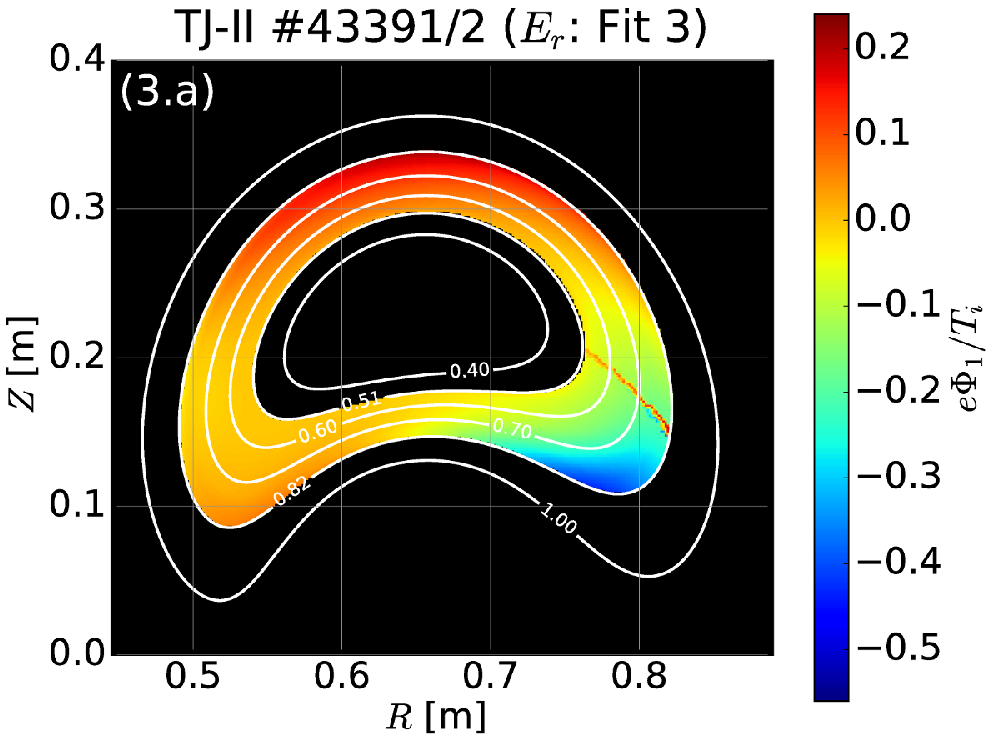}\\
  \includegraphics[height=3.95cm]{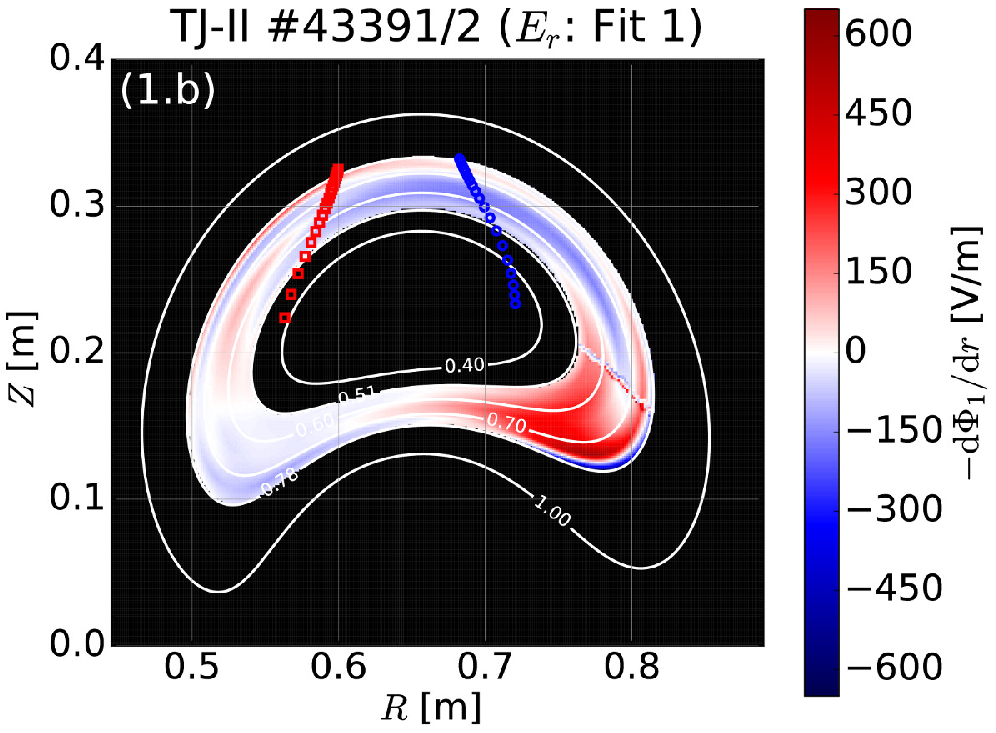}
  \includegraphics[height=3.95cm]{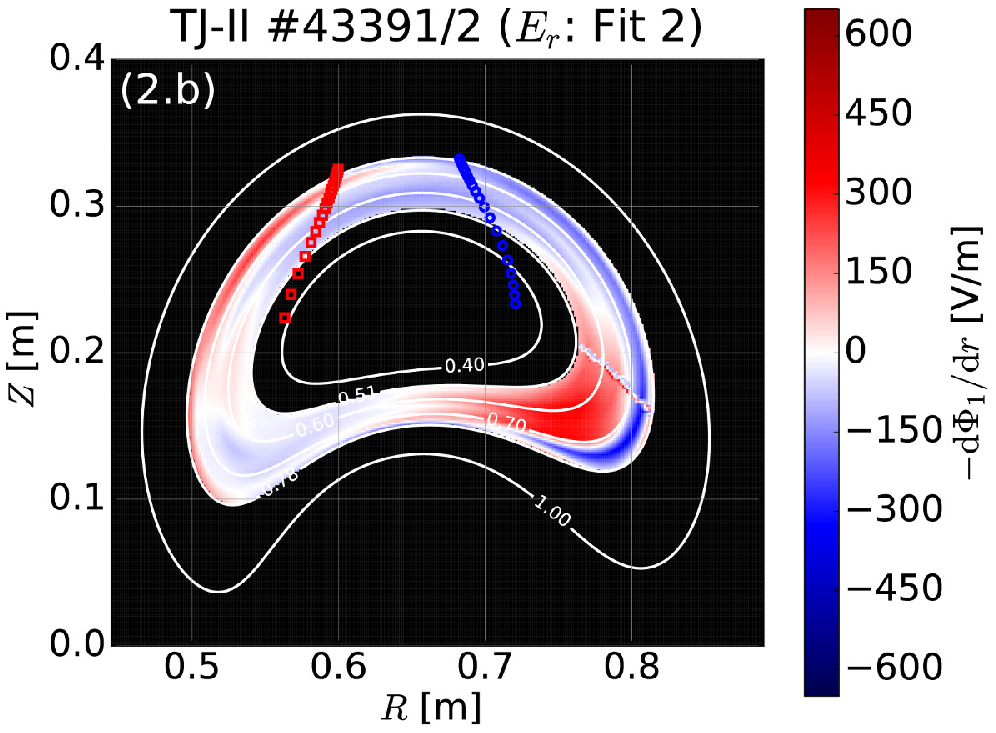}
  \includegraphics[height=3.95cm]{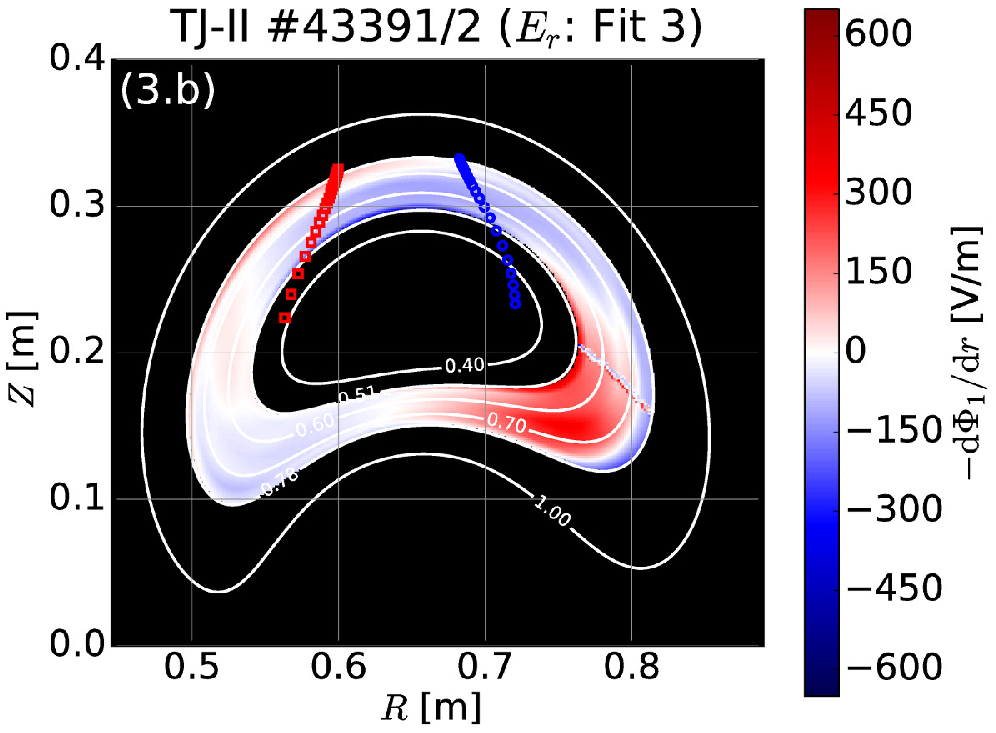}\\  
  \includegraphics[height=5.1cm]{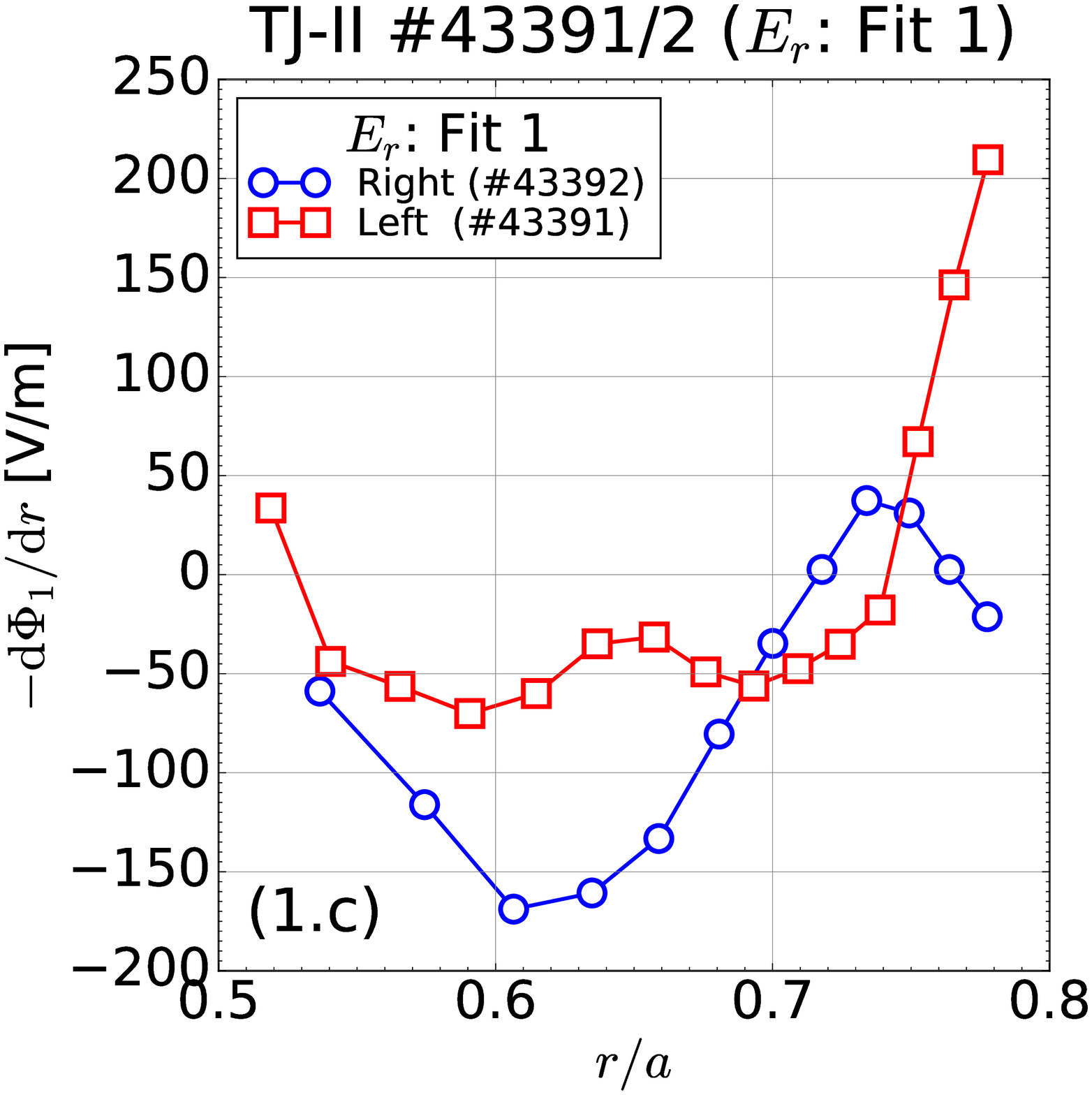}
  \includegraphics[height=5.1cm]{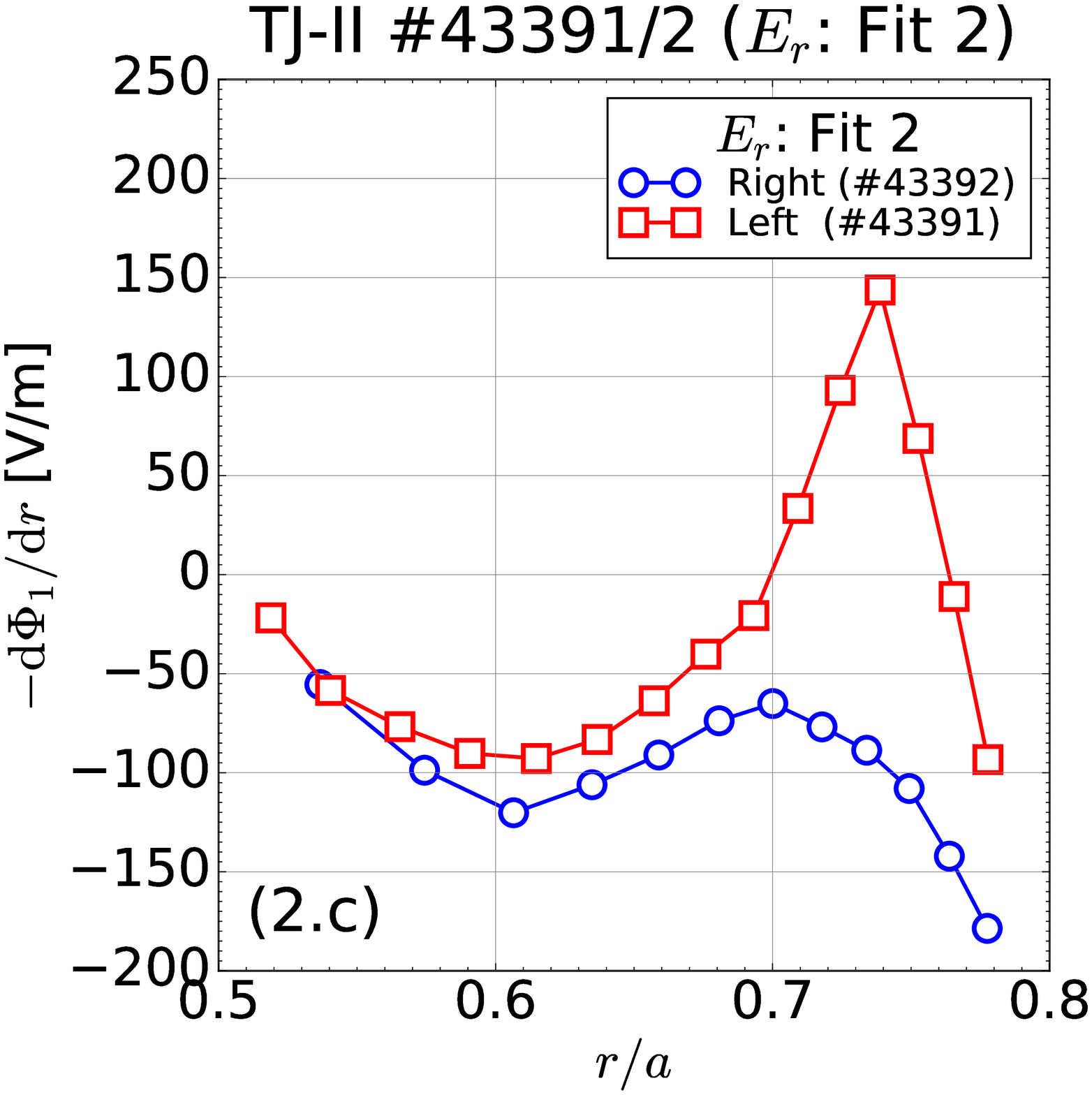}
  \includegraphics[height=5.1cm]{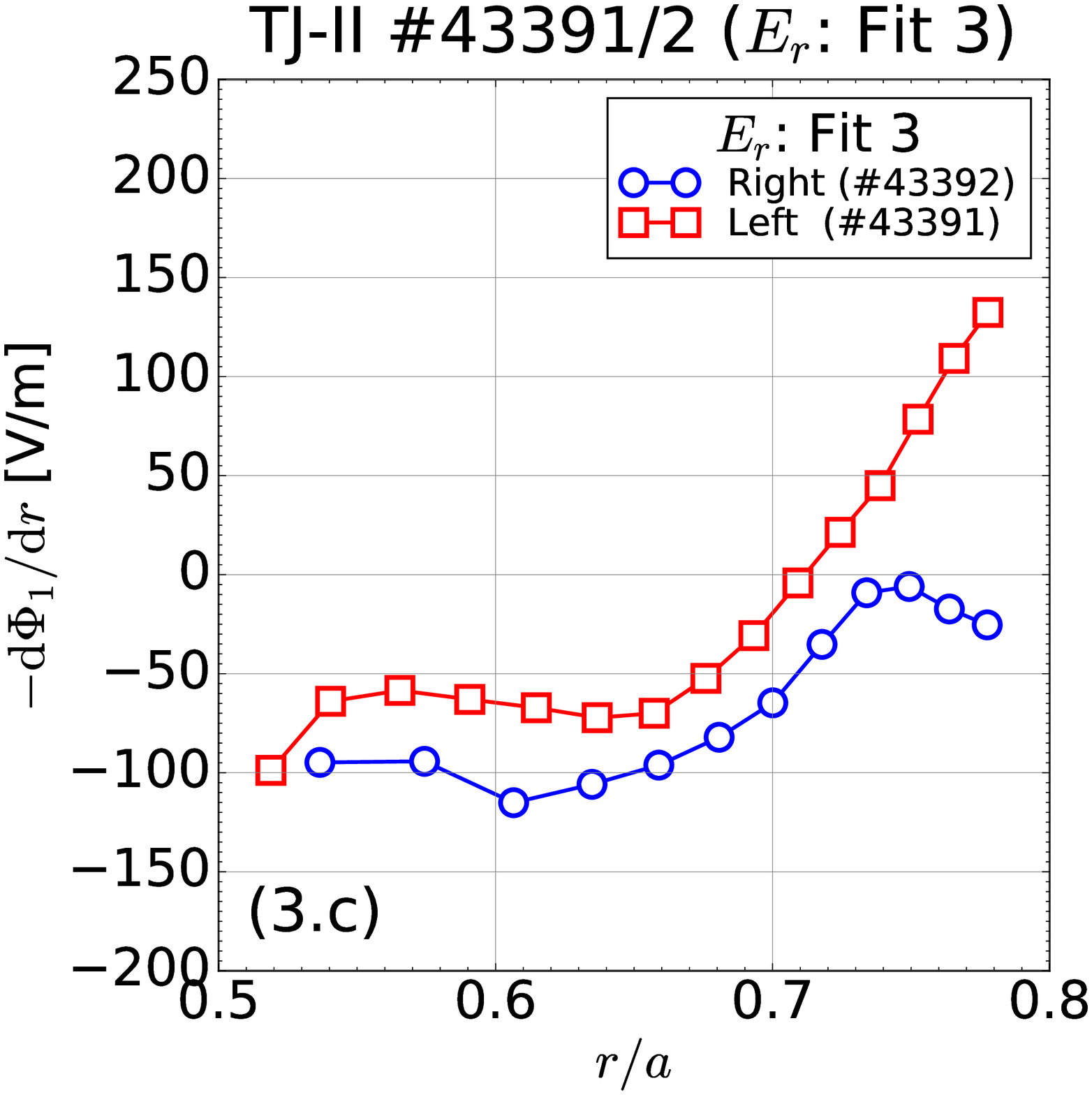}
  \caption{For the electron root conditions TJ-II plasmas: (1-3.a) potential variation normalized to the ion 
temperature $e\Phi_1/T_{i}$ at the Doppler reflectometry probing plane in the range of simulated 
radii, from left to right for the input $E_{r}$ denoted as fit 1 to 3 in \ref{fig:profs_er}(b); (1-3.b) over the same plane, first order radial electric field $-\text{d}\Phi_1/\text{d} r$, together
with the specific positions of measurement on the left and right DR probing regions, 
estimated with ray tracing for the three $E_{r}$ fits considered; (1-3.c) value of 
$-\text{d}\Phi_1/\text{d} r$ at those positions where, as before, red squares and 
right blue circles correspond
to the estimations along the left and right measurement position respectively.}
\label{fig:tj2elec}
\end{figure}

The radial electric field was obtained in the first discharge 
of each pair (this is for the shots $\#43387$ and $\#43391$) on the left side of the 
DR measurement plane.
For the second of the discharges of each pair (this is for shots $\#43388$ and $\#43392$) 
the DR beam was launched to measure the radial electric field on the right side.
It is worth recalling that
the radial electric field provided by the Doppler reflectometer, $E_{r}^{\text{DR}}$,
is obtained from the measured plasma 
background perpendicular flow $u_{\bot}$ and relates to it as $E_{r}^{\text{DR}}=u_{\bot}B$ 
($B$ the modulus of the local magnetic field at the point where the beam 
is reflected).
Assuming the phase velocity of density fluctuations much smaller than 
the $E\times B$ flow velocity, $\mathbf{v}_{E0}=E_{r}\nabla r\times \mathbf{B}/B^2$, 
$u_{\bot}$ is assumed to be equal to the to the latter. Typically the value provided $E_{r}^{\text{DR}}$,
is that of the local radial electric field, which carries with the local dependence of the 
flux expansion term $\nabla r$. This term is comparable in the two plasma regions 
the system can access, and cannot lead to large differences in the local 
radial electric field. But, since the present work focuses on the 
different value of the radial electric field at points located over the same flux surface, 
the modulus of the flux expansion term has been divided out from the experimental 
$E_{r}^{\text{DR}}$ in order to work with, strictly speaking, the supposedly flux function quantity $E_{r}$.
This is
indeed the quantity neoclassical codes require as input.
The radial electric field is represented as a function of the normalized 
effective radius $r/a$ in fig. 
\ref{fig:profs_er} (left) for the ion root discharges and \ref{fig:profs_er} (right) for the electron
root discharges. 
The points with errorbars show the experimental data, 
and the solid lines correspond to different fitted curves used in the EUTERPE simulations 
discussed below.
The values obtained at the left side of the plane of measurement
are represented with red open squares while those taken at the right side are represented by 
blue open circles.
As it is observed in fig. \ref{fig:profs_er} (left), for the ion-root plasmas (shots $\#43387/8$) 
the difference between the radial electric field measured at each side is small, 
in all the accessible radial domain. Only around 
$r/a=0.6$ a slight separation between them can be appreciated. 
On the contrary, under electron root conditions (shots $\#43391/2$),
see fig. \ref{fig:profs_er} (right), the measured radial electric field
is appreciably larger on the right side than than on the left side
on a wide portion of the accessed radial range. In the interval $r/a=0.6-0.8$ discrepancies
of up to $1$ kV/m can be observed. In the numerical analysis that follows we try to 
quantify to what extent the radial dependency of the potential $\Phi_1$ 
can introduce corrections in the total radial electric field 
through the term $-\Phi_1'=-\text{d}\Phi_1/\text{d}r$.\\ 

For the numerical simulations different fitting curves 
for the input ambipolar electric field have been considered. 
They are depicted with solid lines in fig.~\ref{fig:profs_er}.
For the ion root scenario only one case has been used while 
for electron root three have been considered, due to the ambiguity in the choice 
of $E_{r}$ given the 
disparate values measured at each measurement region. One of the curves considers
the data measured on the left side of the probing plane (``fit 1''), 
another curve the data measured on the right side of the plane (``fit 2'')
and a third one the mean value of the previous two (``fit 3'').\\

The numerical results for the ion root case are shown in figs.{\ref{fig:tj2ion}(a) to (c), where 
the following quantities are represented: (a) the potential variation $\Phi_1$ in
a corona of the measurement plane that covers approximately the same radial range as
the experimental data; (b) the radial electric field term $-\Phi_1'$ 
resulting from the potential represented in the previous figure. The measurement positions on
the right and left regions are indicated with read and blue points 
(these positions have been obtained with the ray tracing code TRUBA \cite{TRUBA_manual}); 
(c) $-\Phi_1'$ at the positions indicated in the previous plot. 
The results estimated on the left regions are indicated in red color, while 
those concerning the right side are indicated in blue. 
Looking at figure (c) one can observe that $-\Phi_1'$
shows its maximum values at the outermost radial positions, 
but there is no numerical difference between the value of $-\Phi_1'$ on the left
and right regions. Only around $r/a=0.6$  
the curves in fig. (c) separate from each other a few tens of V/m, which does not represent 
a large difference compared to the value of the ambipolar electric field 
at that position $E_{r}~-600$ V/m. In that sense the numerical
results agree relatively well with the experiment.\\

For the electron-root plasmas the same (a) to (c) plots are represented from 
top to bottom in the set of figs. \ref{fig:tj2elec}, for each of the 
input $E_{r}$ considered for EUTERPE in a different column. Looking at the figs. (1-3.c), in contrast to 
the ion-root case, a more appreciable difference than for the ion root plasmas is observed
between the results for the left and right sides. 
In the three cases 
the correction term $-\Phi_1'$ would make the total radial electric field 
larger on the left side than on the right side, as the curve of $\Phi_1'$ 
indicating the left side values is situated almost at all radii 
above the curve indicating the values on the right region.
along the left array of measurement positions than along the right array. 
The difference between the results with different input $E_{r}$
are given only on the location where the
maximum differences on $-\Phi_1'$ are found. Considering the fit 1, the 
difference reaches up to values of around 200 V/m, and these take place in 
the interval $r/a=0.6-0.7$ and the outermost radial region. 
For fit 2 differences of up to around 250 V/m, larger
on the left than on the right side, are observed at around $r/a\sim 0.75$; 
and finally fit 3 leads to differences that 
only show up at the outermost represented radii, reaching values of around $200$ V/m. 
The numerical difference for the three cases considered are neither as large as those 
found in the DR measurements shown in fig. \ref{fig:profs_er} (right) nor the sign 
coincides numerically and experimentally. In the simulations the radial electric field becomes
larger on the left than on the right probing regions, while in the experiments the opposite
happens. However, it is worth noting that out of the measurement point arrays, following 
a given flux surface contour over the probing plane 
See for instance the reddish areas at the bottom right part of the DR section and 
the top blue areas the contour $r/a=0.5$ passes through in fig.~\ref{fig:tj2elec} (1.a). 
These large deviations cast doubts about the
applicability of the trajectories, eq.~(\ref{eq:dotR_sim})-(\ref{eq:dotmu_sim}) assumed in our simulations, since 
all terms related to $\Phi_1'$ are not present. 
The correction to the total radial electric field arising from $-\Phi_1'$ 
represents in these electron root TJ-II plasmas a significant fraction compared to the
input ambipolar electric field. This is especially the case 
of the lowest $E_r$ fit 1. This 
limits our conclusion quantitatively, and reduces it to the statement that in TJ-II 
electron root plasmas the size of $\Phi_1$ and related contribution to the total radial 
electric field $\Phi_1'$ can become locally a non-negligible fraction of $\Phi_0$ and $E_{r}$, respectively.

\section{Potential variations in W7-X: CERC plasmas and effect of kinetic electrons}
\label{sec:w7xop11}

\begin{figure}
\centering
\includegraphics[width=0.48\textwidth]{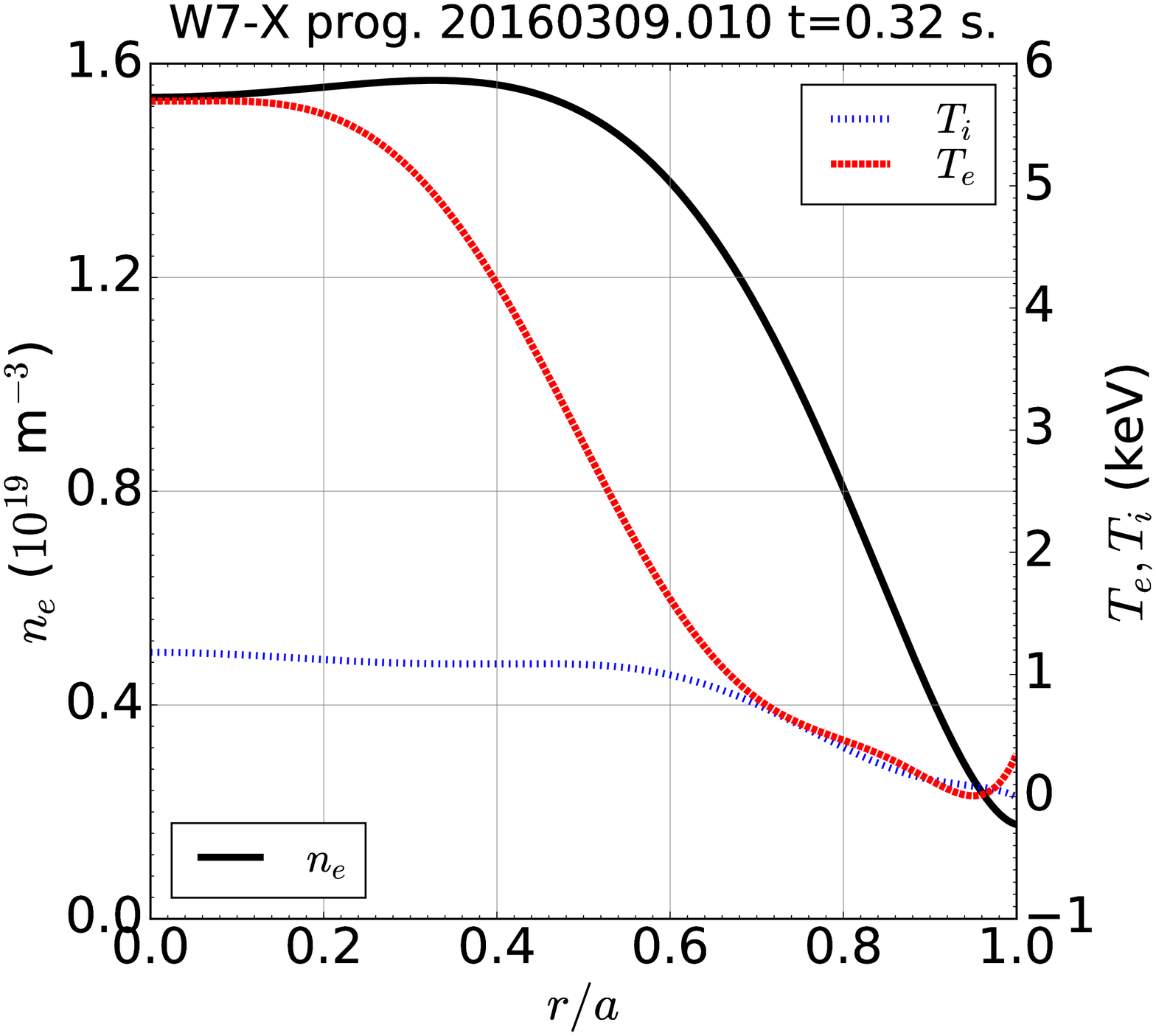}
\includegraphics[width=0.48\textwidth]{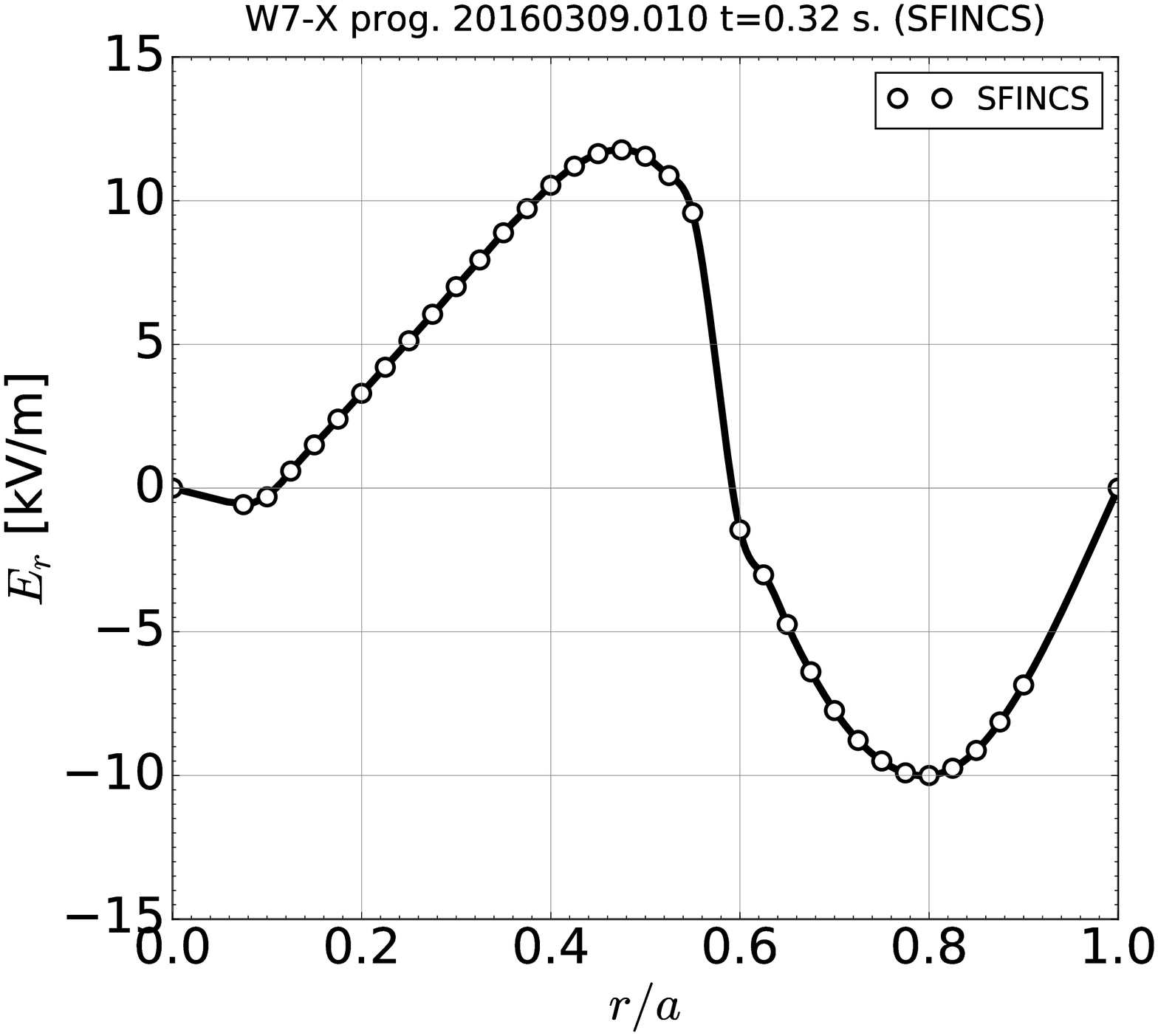}
\caption{(a) Electron density ($n_e$, solid black line), electron temperature ($T_{e}$, dashed red line) and ion temperature ($T_{i}$, dotted blue line) 
considered for the EUTERPE simulations based on those of W7-X programme 20160309.010 at $t=0.32$ ms measured with the
Thomson Scattering ($n_{e}$ and $T_{e}$) and the XICS ($T_{i}$) systems. (b) ambipolar radial electric field 
obtained with the SFINCS code (dots) considering the profiles on the left, and the curve used
as input for EUTERPE.}
\label{fig:w7xop11_profiles}
\end{figure}

\begin{figure}
\centering
\centering
\includegraphics[width=0.48\textwidth]{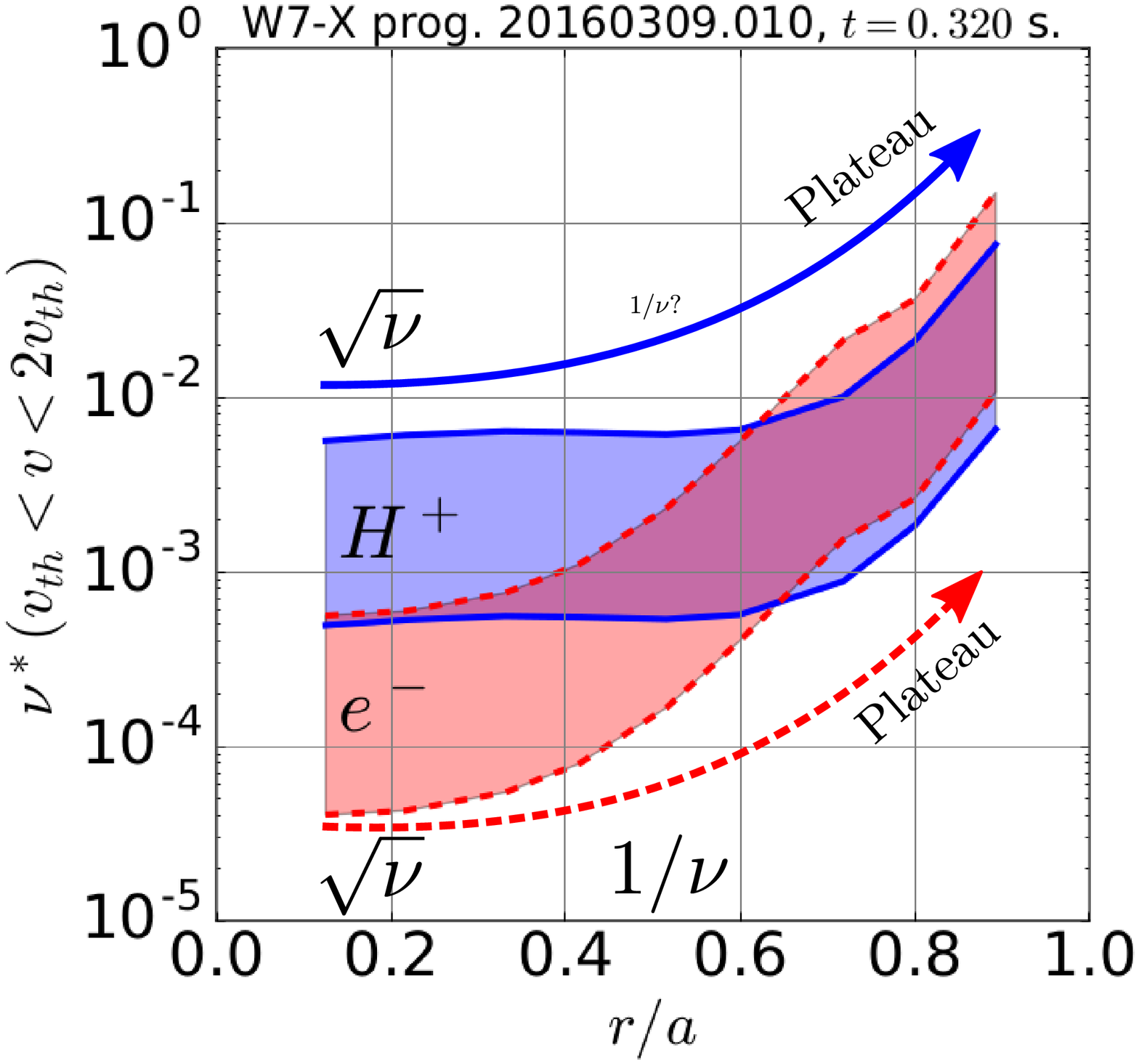}
\includegraphics[width=0.48\textwidth]{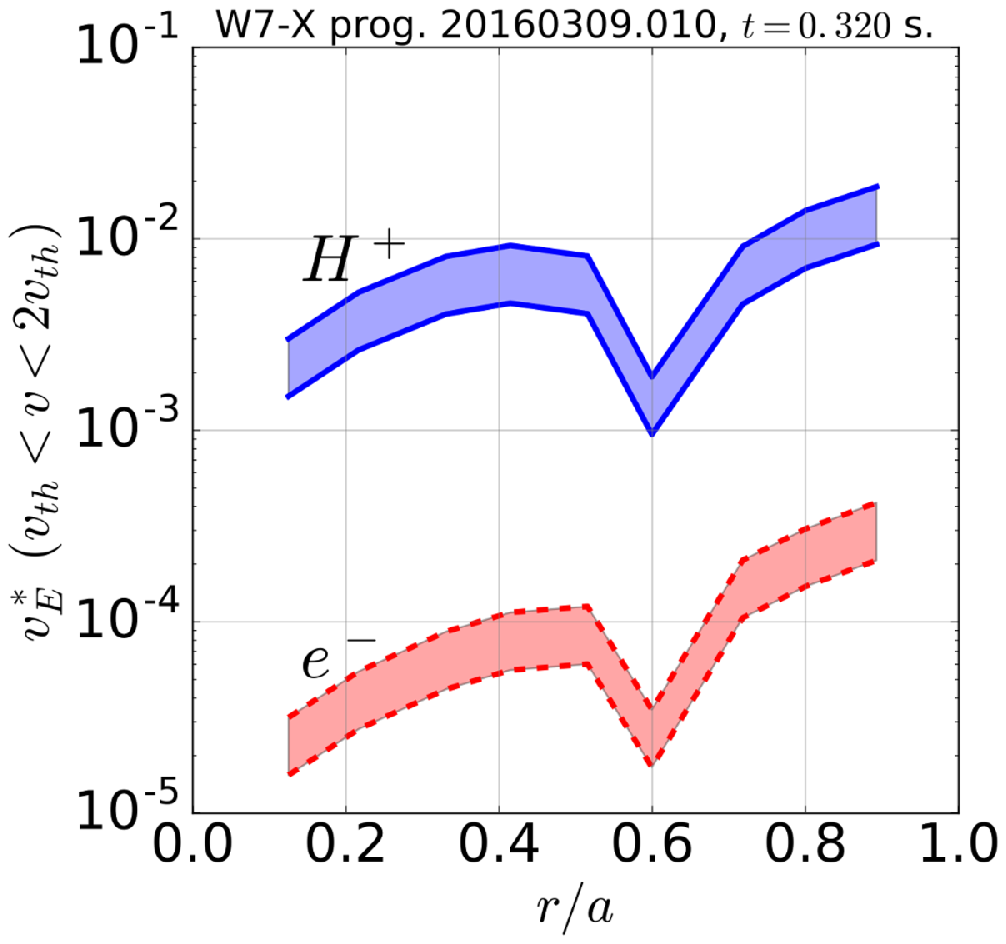}
\caption{Left: normalized collision frequency as a function of the normalized effective
radius for the electrons (red shadowed area) 
and main ions (blue shadowed area) with velocities in the range of one and two thermal 
velocities considering the profiles of fig. \ref{fig:w7xop11_profiles}(a). 
Right: normalized $E_{r}\times B$ velocity for electrons (red shadowed area) 
and main ions (blue shadowed area) with velocities in the range of one and two thermal 
velocities considering the profiles of fig. \ref{fig:w7xop11_profiles}(a) and (b).}
\label{fig:w7xop11_regimes}
\end{figure}

Potential variations have so far been estimated  
small in W7-X plasmas and its impact on impurity transport negligible. However, these conclusions, drawn from the results presented in
refs.~\cite{Regana_nf_57_056004_2017, Mollen_submitted_2018}, cover still a very narrow parameter and configuration 
window of W7-X. In particular all plasmas studied in those references are 
ion root plasmas foreseen during the future W7-X operation phase OP2. 
The calculations were performed for one of the W7-X configurations with lowest neoclassical transport, whose low effective ripple, 
the target figure of merit for the neoclassical optimization and design of W7-X, is lower
than in the configurations for which most of the experiments have been performed.
The case studied in this work widens the parameter window considering CERC plasma parameters from the 
operation phase OP1.1 \cite{Wolf_nf_57_102020_2017}, in particular the physics programme 20160309.010 at the time $t=0.320$. 
The radial profiles based on that programme and instant, used for the simulations discussed below 
are represented in fig. \ref{fig:w7xop11_profiles} (left). The represented profiles are fitted to the Thomson 
Scattering system \cite{Pasch_rsi_87_2016} data for the electron density $n_{e}$ and temperature $T_{e}$, while 
the bulk ion density $T_{i}$ considers the XICS \cite{Langenberg_NF_57_086013_2017} experimental data. 
The radial profile of $E_r$ used as EUTERPE input has been provided by the SFINCS code and is
represented in fig. \ref{fig:w7xop11_profiles} (right).

The reasons for choosing this plasma are the following. On the one hand, 
it is an example of CERC plasma \cite{Yokoyama_nf_47_9_1213_2007} where a root transition takes place. $E_{r}$ is positive 
(electron root) at the inner core and negative (ion root) at the outer part of the core and edge. 
This feature is interesting since, as pointed out in \cite{Pedrosa_nf_55_2015}, under ion root conditions the thermodynamic force 
related to the ambipolar radial electric field opposes to the density and temperature gradients, while in electron root 
all thermodynamic forces have in general for the ions (except deeply hollow profiles, which is not the case here) the same sign. This leads to
a larger source term in the drift kinetic equation that forces the 
perturbed part of the distribution function of the different species $f_{1a}$ to be larger. 
Since the lack of quasi-neutrality among the 
charge density related to these pieces of the distribution function is what gives rise to the 
potential $\Phi_1$, 
this reasoning should lead to expect roughly larger $\Phi_1$ too.
In addition, the change in the direction of the $E\times B$ precession from electron to 
ion root should introduce appreciable changes on the phase of the potential. 
These two statements can be checked by comparing 
how $\Phi_1$ looks on each side of the radial electric field root transition.\\

\begin{figure}
\centering
\centering
\includegraphics[width=0.9\textwidth]{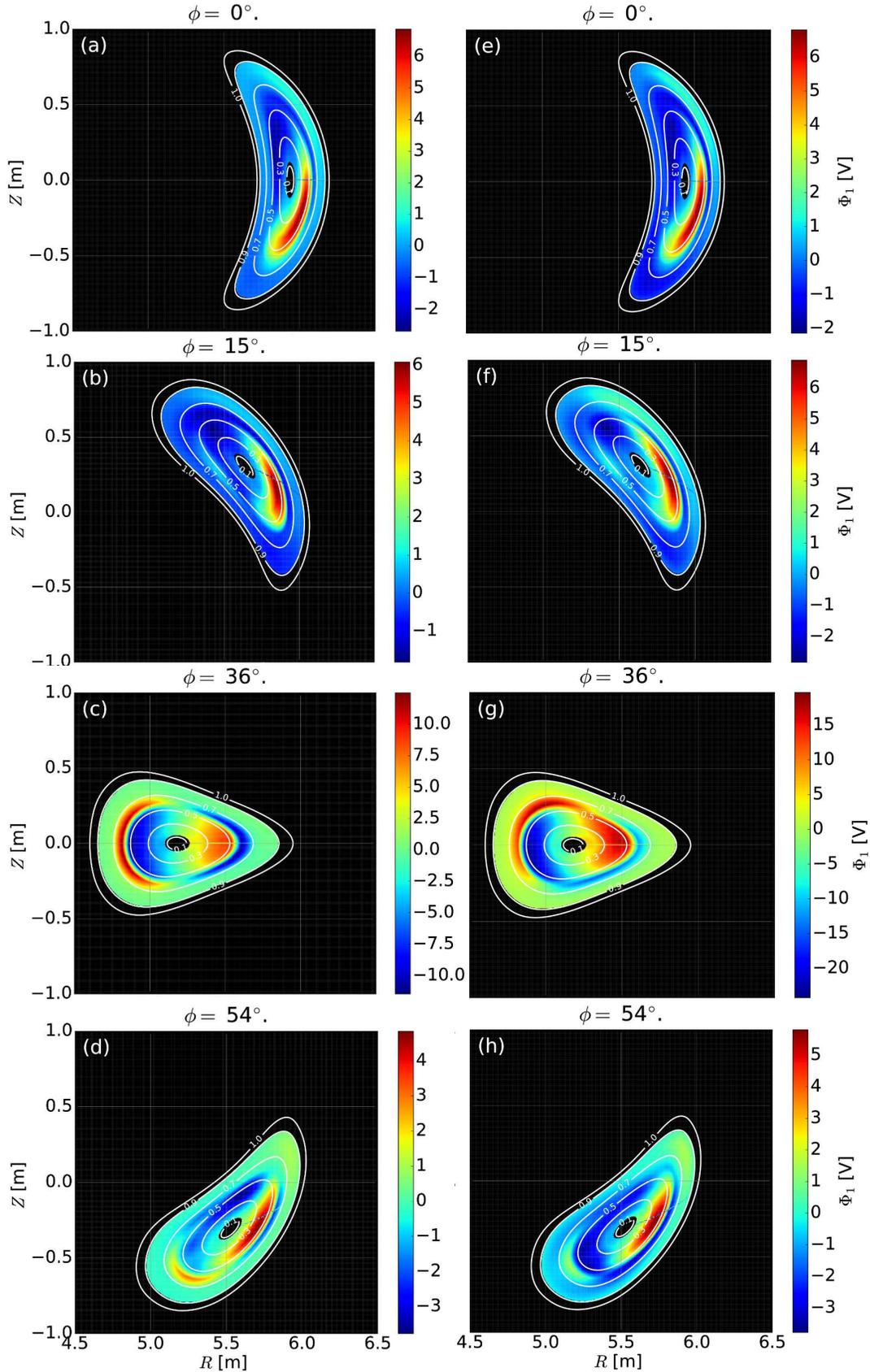}
\caption{For the simulations with adiabatic electrons (left column) and 
kinetic electrons (right column), from top to bottom: 
calculated potential in W7-X at the toroidal planes $\zeta=0, 15, 36$ and $54^\circ$. \blue{Note the different 
color scales on the left and right plots, employed to appreciate the changes in the shape of $\Phi_1$ between considering 
adiabatic or kinetic electrons.}}
\label{fig:w7xop11_phi1_comp}
\end{figure}

\begin{figure}
\centering
\includegraphics[width=0.48\textwidth]{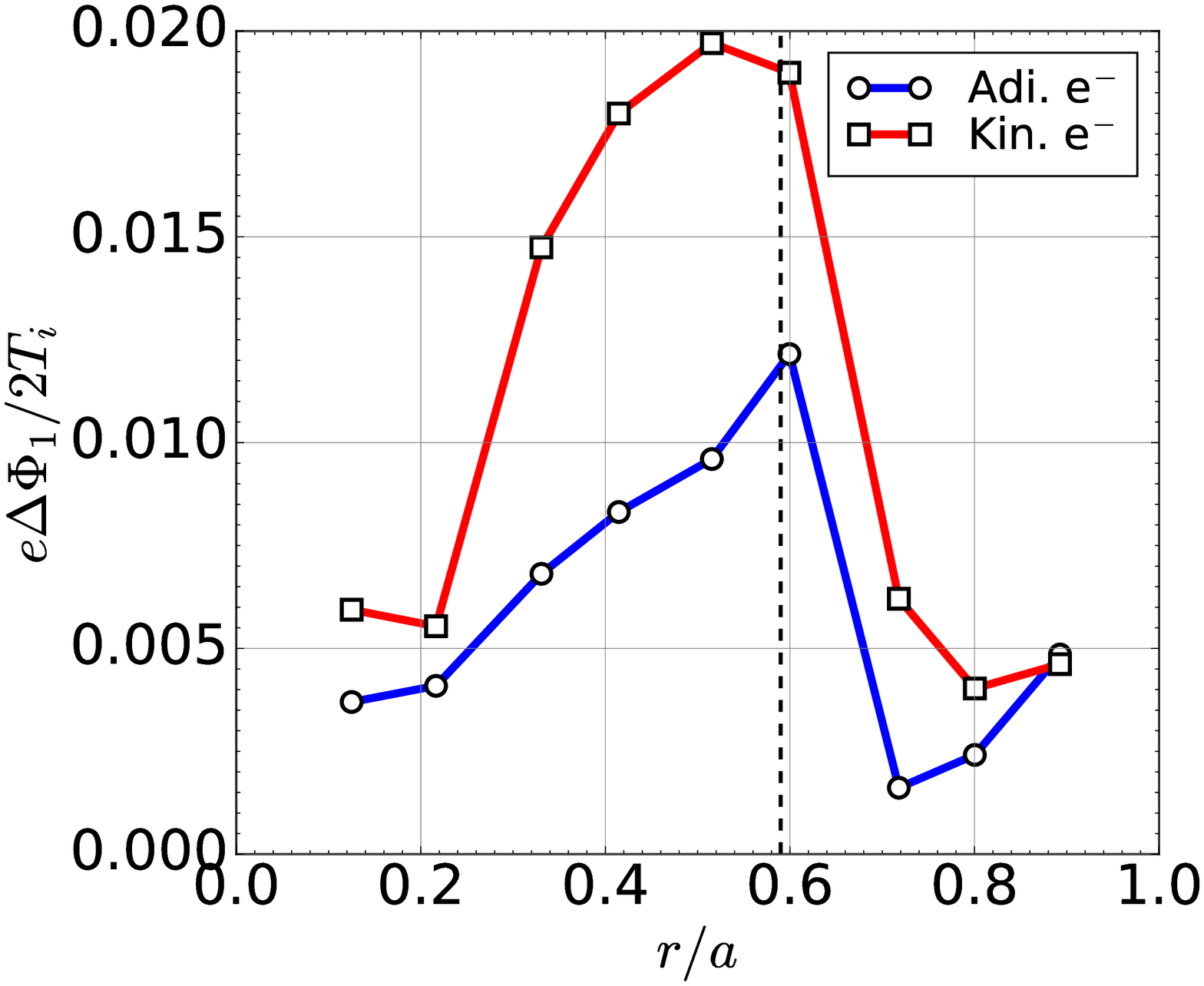}
\caption{Maximum difference of the potential $\Delta\Phi_1=\Phi_{1}^{\text{max}}-\Phi_{1}^{\text{min}}$ 
on each of the simulated flux surfaces 
normalized to the ion temperature $T_{i}$ as a function of the normalized effective radius $r/a$
for the calculations with adiabatic electrons (circles connected with blue segments) and kinetic electrons (squares
connected with red segments). \blue{The dashed vertical line indicates the radial position 
where $E_{r}=0$. On the left and right of this line the input radial electric field is positive 
and negative, respectively.}}
\label{fig:w7xop11_deltaPhi1_comp}
\end{figure}

Furthermore the fact that the temperature of the electrons
is significantly higher than that of the ions leads to a situation where the electron contribution ECH
to $\Phi_1$ eventually may become important. Note that in ref.~\cite{Regana_nf_57_056004_2017}
the electrons are considered adiabiatic, based on the condition $T_{i}\sim T_{e}$ and the higher density of
the plasmas there, and thus the electron contribution to $\Phi_1$ is neglected.
In order to know whether the electrons may contribute to $\Phi_1$, let us recall first that for a given magnetic configuration and 
for one single kinetic energy or velocity $v$, 
the parameters to find in which collisional regime each species is,
are the normalized $E_{r}\times B$ drift velocity $v_{E}^{*}=E_{r}/v B_{0}$ and the 
normalized collision frequency $\nu^{*} = R_{0}\nu/(\iota v)$. See for instance ref.~\cite{Beidler_nf_51_076001_2011},
where several configurations are considered and the main thermal transport 
matrix coefficients are represented as a function of $\nu^{*}$ for different values of $v_{E}^{*}$.
In particular, in the scalings depicted for the normalized transport matrix coefficient $D_{11}^{*}$, 
helpful visual references of the collisionality interval at which the $1/\nu$ scaling begins 
and when transits to the $\sqrt{\nu}$ regime are found.
This so-called mono-energetic view is somewhat limited since
the Maxwellian velocity distribution function covers a range of velocities and not just one.
In fig. \ref{fig:w7xop11_regimes} the range of $\nu^{*}$ and $v_{E}^{*}$ values as a function of $r/a$ are represented for electrons
and ions (H$^{+}$) with velocity between $v=v_{th}$ and $v=2 v_{th}$, considering the plasma parameters of 
fig. \ref{fig:w7xop11_profiles}. Looking at the values of 
$\nu^{*}$ and $v_{E}^{*}$ for the ion parameters and comparing with the scanned ranges in ref.~\cite{Beidler_nf_51_076001_2011} for W7-X, 
one can conclude that the ions should mostly be in the $\sqrt{\nu}$ regime at the 
innermost radial positions and in the plateau regime at the edge, 
passing through a practically inexistent $1/\nu$ regime. Regarding the electrons, their much lower 
normalized collisionality compared to that of the ions at the core make them mainly reside in 
the $\sqrt{\nu}$ regime in that region as well. They also exhibit $\nu^{*}$ values at the edge
characteristic of the plateau regime but, contrarily to the ions, the much lower 
$v_{E}^{*}$ binds them to pass through a more robust and wider (in collisionality) $1/\nu$ regime in between. This 
consequently should make the electrons to necessarily be in a deep $1/\nu$ regime on a radially wide region of the core. 
Finally, since the perturbed part
of the distribution function (and consequently the perturbed part of the density entering in the equation for the 
potential variation) scales in the $1/\nu$ regime with $\rho^{*}/\nu^{*}$ while in the $\sqrt{\nu}$ regime
is independent of $\rho^{*}$ and $\nu^{*}$ (with $\rho^{*}$ the normalized Larmor radius to the stellarator size)
\cite{Calvo_ppcf_59_055014_2017, Calvo_jpp_inprogress_2018}, 
the core of these plasmas are particularly favorable to show differences between considering kinetic or adiabatic electrons
in the calculations of $\Phi_1$. This is the numerical comparison 
presented and discussed in the following paragraph.\\

The calculations of $\Phi_1$ have been performed for nine radial positions, approximately 
separated between each other $\Delta r/a = 0.1$. These 
radii are $r/a=\lbrace 0.12, 0.22, 0.33, 0.41, 0.51, 0.60, 0.72, 0.80, 0.90\rbrace$. 
The figures of the potential on the toroidal planes have been obtained by interpolation
using the value over the simulated flux surfaces.
In fig.~\ref{fig:w7xop11_phi1_comp} the potential variation 
is represented, from top to bottom, for the toroidal planes $\phi=0^\circ, 15^\circ, 36^\circ$ and $54^\circ$. 
The first of these toroidal planes has the practical interest that a Doppler and a correlation reflectometer 
probe that plane in order to characterize the experimental radial electric field. 
At $\phi=15^{\circ}$ a second Doppler reflectometer is also installed. 
The other two planes have been considered since the distance 
between them in $\phi$ is one fourth of a the machine period, the first of them being the
frequently represented triangular plane where other essential diagnostics for impurity 
transport look at, like the soft-X rays Multi-Camera Tomography System (XMCTS) \cite{Brandt_fed_123_2017} 
or the bolometry cameras \cite{Zhang_rsi_81_2010}. The difference between the figures on the left, with
labels (a)-(d), and on the right, with labels (e)-(h),
 is that while the former show the results assuming adiabatic electrons, 
the latter do it for the cases considering kinetic electrons. 
First of all, note that the range in the color scale changes from plot to plot, in order to make appreciable the changes 
in the shape of the potential, that keeping the same scale for all cross sections would not allow to appreciate. 
Looking at those color scales and their ranges, it can be seen that the largest 
$\Phi_1$ values are very localized on the triangular plane, where they become much larger than on the other planes.
Second, the size of the potential for the case with kinetic electrons is roughly up to twice as large as 
when the electrons are assumed adiabatic. This is evident on the triangular plane while on the other
the difference is not remarkable.
Looking at the potential at the triangular plane, it is also observed that 
the shape experiences appreciable changes when the electrons are considered 
as a kinetic species compared to the case with adiabatic electrons. 
In particular, the negative values of $\Phi_1$, that in the case with adiabatic electrons
\ref{fig:w7xop11_phi1_comp}(c) are located 
on the low field side (LFS) and below
the equatorial plane, are displaced towards the high field side (HFS) when electrons are kinetic \ref{fig:w7xop11_phi1_comp}(g). 
This is also compatible with what is known about the symmetry properties of $\Phi_1$ \cite{Alonso_spineto_2017}. 
When only the contribution from the ions is considered, since they must be mostly in the $\sqrt{\nu}$
regime, $\Phi_1$ must necessarily have cosine components dominating its spectrum, leading to the 
clear in-out asymmetry that fig. \ref{fig:w7xop11_phi1_comp}(c) illustrates. When kinetic electrons are considered,
since they must, as we have hypothesized, add their contribution from the $1/\nu$ 
regime, the consequent introduction of sine component leads that in-out asymmetry to blur as
\ref{fig:w7xop11_phi1_comp}(g) shows.
Other changes in the shape are observed in other planes, although not as clear as 
on the triangular plane.\\

\blue{Other features can more clearly be observed in fig. \ref{fig:w7xop11_deltaPhi1_comp}, where  
the maximum normalized potential difference ($\Delta\Phi_1=(\Phi_1^{\mathrm{max}}-\Phi_1^{\mathrm{min}})/2T_{i}$) is 
represented. 
The results are shown for both calculations, with 
adiabatic electrons and with kinetic electrons. 
Roughly speaking the potential variation size is shown to be considerably larger in the 
portion of the plasma in electron root than in that under ion root. 
In addition, a much larger contribution of the kinetic 
electron response is observed in the first of these regions than in the second. 
However, the point located in ion root immediately 
after the root change (i.e. at $r/a=0.6$) exhibits a large value as well.
A vertical line represents the exact position where the root change takes place in the
input $E_{r}$.
At that point the ambipolar electric field is rather low $E_{r}=-1.38$ kV/m, compared
to the value at the other positions in ion root where $|E_{r}|>7$ kV/m. This low value of $E_{r}$ 
can be the cause of adding a large contribution to $\Phi_1$ from ions in the $1/\nu$ regime.}
\blue{Another interesting feature results from the large variations 
at each side of the root transition together with the abrupt change of its phase.
To appreciate this one can look at the triangular plane represented in 
fig. \ref{fig:w7xop11_phi1_comp}(g) for the calculation including kinetic electrons. 
This change is present at almost
any poloidal position in the vicinity of that radius and is given 
in a relatively narrow region (the two radii simulated immediately before and after the root change are separated 
by $\Delta r/a=0.09$). It is then natural to ask whether this can introduce some important
contribution to the radial electric field. This correction, $-\Phi_1'$, 
is represented at the triangular plane, considering kinetic electrons, in fig. \ref{fig:er1triangular}.
Moderate values of a few hundreds of V/m are present on that cross section but near to the 
root change the value is considerably larger, reaching around 1 kV/m, both positive and negative. 
In our characteristic trajectories, see eqs.~(\ref{eq:dotR_sim})-(\ref{eq:dotmu_sim}), the 
$E\times B$ drift related to this component of the radial electric field is not kept
to lowest order. This is applicable since, as it happens at almost all positions, 
$E_{r}$ is substantially larger than the represented $-\Phi_1'$. However, it becomes of the same order at 
the innermost simulated radius under ion root conditions (where as above-mentioned $E_{r}$ was $-1.38$ kV/m). 
Then, to this respect the calculations on that specific position should be taken, as 
well as the conclusions drawn from it, cautiously.}

\begin{figure}
\centering
\centering
\includegraphics[width=0.65\textwidth]{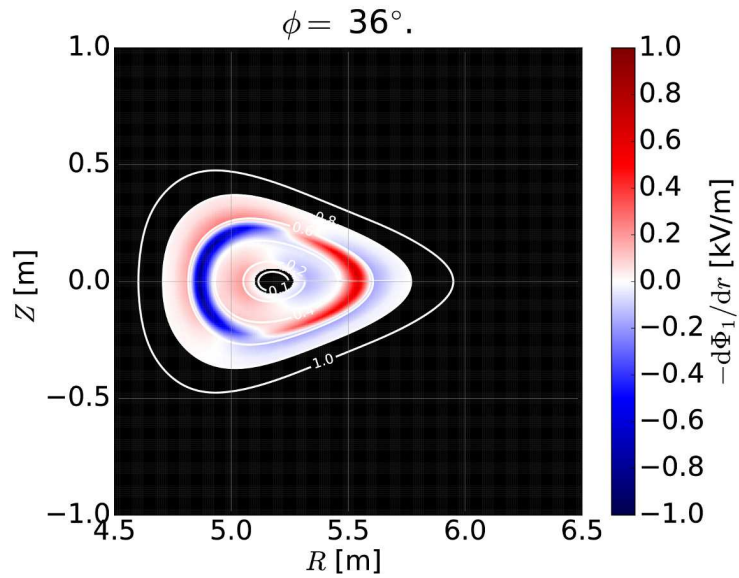}
\caption{First order electric field contribution $-\text{d}\Phi_1/\text{d}r$ obtained at the toroidal 
plane $\phi=36^\circ$, considering the corresponding potential $\Phi_1$ for the case with kinetic electrons.}
\label{fig:er1triangular}
\end{figure}

\section{Conclusions}
\label{sec:conclusions}

The present work has addressed the calculation of the neoclassical potential variation, 
with the emphasis on electron-root plasmas. The standard configuration 
for TJ-II and a high mirror configuration of W7-X have been used, 
considering plasma parameters of discharges from their recent experimental campaigns.\\

In TJ-II, the Doppler Reflectometry radial electric field measurements and, 
in particular, the strong difference of its value at different points over 
the same flux surface, has motivated looking into
the radial dependence of $\Phi_1$ and investigate to what extent the 
term $-\Phi_1'$ contributes to the total radial electric field.
What has been found by numerical simulations agrees qualitatively with the experimental 
results. The difference in the total electric field that the potential variations
can make is large in the electron root cases, although still a non-negligible factor smaller than the experimental one.  
On the other hand this correction is practically not present in the ion root plasmas, 
both numerically and experimentally. 
These conclusions are drawn from the comparison made at the specific measurement positions of the 
Doppler reflectometry system 
over the same flux surface. 
Out of these locations $-\Phi_1'$ is found large both under ion and 
electron root conditions. This points to some concern about the 
applicability of the characteristic trajectories of the simulated particles, since
terms containing $-\Phi_1'$ are neglected based on its size compared to $-\Phi_0'$,
although \textit{a posteriori} the former is found not too small compared to the latter.
This and the possibility that the kinetic electrons could introduce a non-negligible
contribution to the potential, as proven in the section by the numerical simulations results for W7-X, 
are possible reasons than may have frustrated a better agreement.

Regarding W7-X we have considered a configuration with significantly larger
effective ripple than the standard configuration analyzed in past works 
\cite{Regana_nf_57_056004_2017}. 
The plasma parameters correspond to a standard CERC plasma from OP1.1. 
The analysis has demonstrated that W7-X can access regimes with potential
variations significantly larger than what has already been reported. In this occasion 
the simulations have been performed with adiabatic and kinetic electrons. 
The comparison between them have shown that the contribution from the kinetic electron
response, when the parameters
are such that they are likely to be deeply in the $1/\nu$ regime, can be significant in the 
size and shape of the potential. This occurred mostly in a broad portion of the plasma in electron root, 
where in addition, the resulting size of $\Phi_1$ was considerably larger than in ion root. 
Other features have been found, like the localization of these large variations 
on the triangular plane of W7-X, and the smaller values near the boundaries of the machine period.
Interestingly, on that triangular plane, at each side of the 
root transition and at the closest radii the potential reaches its maximum values. This, together
with the fact that the phase is the opposite on one side and the other of the root change, 
gives rise to a large radial electric field corrections arising from $\Phi_1'$. 

\section*{Acknowledgements}
\label{sec:acknowledgements}
This work has been carried out within the framework of the EUROfusion Consortium and has received funding 
from the Euratom research and training programme 2014-2018 under grant agreement No 633053. The views and 
opinions expressed herein do not necessarily reflect those of the European Commission.\\
This research was supported in part by grants ENE2015-70142-P and FIS2017-88892-P, Ministerio de Econom\'ia y Competitividad, Spain.

\section*{References}
\label{Bibliography}
\bibliographystyle{unsrt}
\bibliography{/home/jose/Dropbox/biblio.bib}

\end{document}